\documentclass[aps,prd,reprint,superscriptaddress,showkeys,showpacs]{revtex4-1}

\usepackage{graphicx} %
\usepackage{amsmath} %
\usepackage{amssymb} %
\usepackage{subfigure} %
\usepackage{float} %
\usepackage{upgreek} %
\usepackage{multirow} %
\usepackage{color} %
\usepackage{lineno} %
\usepackage{hyperref} %
\hypersetup{colorlinks=true, urlcolor=blue, linkcolor=blue, citecolor=red} %
\begin{document}

%%% Some user definitions for convenience

\newcommand{\dm}{DM-Ice17}

\newcommand{\iso}[2]{\ensuremath{^{#2}\mathrm{#1}}}

\newcommand{\startdate}{July 2011}
\newcommand{\stopdate}{June 2013}

\newcommand{\dma}{Det-1}
\newcommand{\dmb}{Det-2}

\newcommand{\black}[1]{{\color{black}{#1}}}
\newcommand{\red}[1]{{\color{Red}{\textbf{#1}}}} % means doulbe check
\newcommand{\blue}[1]{{\color{Blue}{\textbf{#1}}}}
\newcommand{\green}[1]{{\color{Green}{\textbf{#1}}}} % means try to re-word
\newcommand{\gray}[1]{{\color{gray}{#1}}}
\newcommand{\orange}[1]{{\color{BurntOrange}{#1}}}

\newcommand{\fw}{\textwidth}
\newcommand{\tqw}{0.75\textwidth}
\newcommand{\hw}{0.48\textwidth} %%% for 2-column view
\newcommand{\qw}{0.23\textwidth}

%%% Referencing
\newcommand{\fig}[1]{Fig.\,\ref{#1}}
\newcommand{\tab}[1]{Table\,\ref{#1}}
\newcommand{\sect}[1]{Sec.\,\ref{#1}}
\newcommand{\chap}[1]{Chap.\,\ref{#1}}
\newcommand{\equ}[1]{Eq.\,\ref{#1}}
\newcommand{\cit}[1]{Ref.\,\cite{#1}}

%%% UNITS
%\renewcommand{\deg}{\,C}
\renewcommand{\deg}{\ensuremath{{^{\circ}}\mathrm{C}}}

\newcommand{\aprox}{\ensuremath{\sim}\,}
\newcommand{\per}{\%}
\newcommand{\kpa}{\,kPa}
\newcommand{\hz}{\,Hz}
\newcommand{\s}{\,sec}
\newcommand{\lt}{\ensuremath{\,<\,}}
\newcommand{\gt}{\ensuremath{\,>\,}}
\newcommand{\plm}{\ensuremath{\,\pm}\,}
\newcommand{\kg}{\,kg}
\newcommand{\kgyr}{\,kg\ensuremath{\cdot}yr}
\newcommand{\gs}{\ensuremath{\,g}}
\newcommand{\diam}{\ensuremath{\varnothing\,}}
\newcommand{\usec}{\,\ensuremath{\upmu\mathrm{s}}}
\newcommand{\ubk}{\,\ensuremath{\upmu}Bq/kg}
\newcommand{\mbk}{\,mBq/kg}

\newcommand{\kev}{\,\ensuremath{\mathrm{keV}}}
\newcommand{\mev}{\,\ensuremath{\mathrm{MeV}}}
\newcommand{\kevee}{\,\ensuremath{\mathrm{keV}_{\mathrm{ee}}}}
\newcommand{\kevr}{\,\ensuremath{\mathrm{keV}_{\mathrm{r}}}}
\newcommand{\dru}{\,\ensuremath{\mathrm{counts/day/keV/kg}}}

\newcommand{\gev}{\ensuremath{{\rm GeV}/{\rm c}^2}}
\newcommand{\smod}{\ensuremath{{S_m}}}
\newcommand{\snaught}{\ensuremath{{S_0}}}
\newcommand{\rzero}{\ensuremath{{R_0}}}
\newcommand{\nzero}{\ensuremath{{N_0}}}

\newcommand{\e}[1]{\ensuremath{\times 10^{#1}}}

\title{First data from \dm}

%% the author list
%%%%%%%%%%%%%%%%%%%%%%%%%%%%%%%%%%%%%%%%%%%%%%%%%%%%%%%%%%%%
%%  Authors and affiliations
%%%%%%%%%%%%%%%%%%%%%%%%%%%%%%%%%%%%%%%%%%%%%%%%%%%%%%%%%%%%
%%  \author{}
%%  \affiliation{}
%%  \email{}
%%%%%%%%%%%%%%%%%%%%%%%%%%%%%%%%%%%%%%%%%%%%%%%%%%%%%%%%%%%%
\author {J.~Cherwinka}
\affiliation {Physical Sciences Laboratory, University of Wisconsin-Madison, Stoughton, WI 53589, USA}
\author {D.~Grant}
\affiliation {Department of Physics, University of Alberta, Edmonton, Alberta, Canada}
\author {F.~Halzen}
\affiliation {Department of Physics and Wisconsin IceCube Particle Astrophysics Center, University of Wisconsin-Madison, Madison, WI 53706, USA}
\author {K.M.~Heeger}
\affiliation {Department of Physics, Yale University, New Haven, CT 06520, USA}
\author {L.~Hsu}
\affiliation {Fermi National Accelerator Laboratory, Batavia, IL 60510, USA}
\author {A.J.F.~Hubbard}
\affiliation {Department of Physics and Wisconsin IceCube Particle Astrophysics Center, University of Wisconsin-Madison, Madison, WI 53706, USA}
\affiliation {Department of Physics, Yale University, New Haven, CT 06520, USA}
\author {A.~Karle}
\affiliation {Department of Physics and Wisconsin IceCube Particle Astrophysics Center, University of Wisconsin-Madison, Madison, WI 53706, USA}
\author {M.~Kauer}
\affiliation {Department of Physics and Wisconsin IceCube Particle Astrophysics Center, University of Wisconsin-Madison, Madison, WI 53706, USA}
\affiliation {Department of Physics, Yale University, New Haven, CT 06520, USA}
\author {V.A.~Kudryavtsev}
\author {C.~Macdonald}
\affiliation {Department of Physics and Astronomy, University of Sheffield, Sheffield, UK}
\author {R.H.~Maruyama}
\email[Corresponding author: ]{reina.maruyama@yale.edu}
\affiliation {Department of Physics, Yale University, New Haven, CT 06520, USA}
\author {S.~Paling}
\affiliation{STFC Boulby Underground Science Facility, Boulby Mine, Cleveland, UK}
\author {W.~Pettus}
\author {Z.P.~Pierpoint}
\author {B.N.~Reilly}
\altaffiliation[Present address: ]{Physics Department, University of Wisconsin-Fox Valley, Menasha, WI, USA}
\affiliation {Department of Physics and Wisconsin IceCube Particle Astrophysics Center, University of Wisconsin-Madison, Madison, WI 53706, USA}
\affiliation {Department of Physics, Yale University, New Haven, CT 06520, USA}
\author {M.~Robinson}
\affiliation {Department of Physics and Astronomy, University of Sheffield, Sheffield, UK}
\author {P.~Sandstrom}
\affiliation {Department of Physics and Wisconsin IceCube Particle Astrophysics Center, University of Wisconsin-Madison, Madison, WI 53706, USA}
\author {N.J.C.~Spooner}
\author {S.~Telfer}
\affiliation {Department of Physics and Astronomy, University of Sheffield, Sheffield, UK}
\author {L.~Yang}
\affiliation {Department of Physics, University of Illinois at Urbana-Champaign, Urbana, IL 61801, USA}
\collaboration{The DM--Ice Collaboration}
\noaffiliation
\date{10 November 2014}

\begin{abstract}
We report the first analysis of background data from \dm, a direct-detection dark matter experiment consisting of 17\,kg of NaI(Tl) target material. It was codeployed with IceCube 2457\,m deep in the South Pole glacial ice in December 2010 and is the first such detector operating in the Southern Hemisphere. The background rate in the 6.5\,--\,8.0\kevee\ region is measured to be 7.9\plm 0.4\dru. This is consistent with the expected background from the detector assemblies with negligible contributions from the surrounding ice.  The successful deployment and operation of \dm\ establishes the South Pole ice as a viable location for future underground, low-background experiments in the Southern Hemisphere. The detector assembly and deployment are described here, as well as the analysis of the \dm\ backgrounds based on data from the first two years of operation after commissioning, \startdate\,--\,\stopdate.


\end{abstract}

\maketitle

%% Start line numbering here
%\linenumbers

%%  MAIN TEXT %%%%%%%%%%%%%%%%%%%%%%%%%%%%%%%%%%%%%%%%%%%%%%%
% !TEX root = dm-ice17_performance.tex

\section{Introduction}
\label{sec:intro}

Astrophysical and cosmological observations suggest that roughly 27\per\ of the Universe is cold dark matter~\cite{Ade:2013zuv}. Although evidence for dark matter has been firmly established~\cite{Bertone:2004pz, Faber:1979pp}, its composition and characteristics remain largely unknown. Weakly interacting massive particles (WIMPs) are theoretically favored because they can be produced in the early universe with the correct abundance to result in the observed relic density~\cite{Steigman:1984ac}. A suite of direct detection experiments is now underway~\cite{Cushman:2013zza} to search for WIMPs through observation of WIMP-nucleon scattering~\cite{Goodman:1984dc, Drukier:1986tm}. 
%Many of these experiments now have the sensitivity to probe the region of parameter space preferred by the Minimal Supersymmetric Standard Model~\cite{ Jungman:1995df}.

We report on the performance of \dm, a NaI(Tl) direct dark matter detector deployed at the South Pole in December 2010. \dm\ is designed to demonstrate the feasibility of operating a remote low-background NaI experiment to directly test the annual modulation of the WIMP-nucleon scattering rate observed by DAMA. The expected annual modulation arises from the motion of Earth around the Sun while the Solar System moves through the dark matter halo of our Galaxy~\cite{Freese:1987wu, Chang:2009yt}. The DAMA/NaI~\cite{Bernabei:2005hj} and DAMA/LIBRA~\cite{Bernabei:2013xsa} experiments, running at the Laboratori Nazionali del Gran Sasso for a combined 14-year period, have measured a consistent annual modulation at 9.3\,$\sigma$, which they attribute to dark matter. More recently the CoGeNT~\cite{Aalseth:2012if, Aalseth:2014eft}, CRESST~\cite{Angloher:2011uu}, and CDMS-II(Si)~\cite{Agnese:2013rvf} experiments have observed events in excess of the known backgrounds in their respective detectors. Under the assumption of elastic scattering of WIMPs, these results are inconsistent with exclusion limits set by several direct detection experiments for both spin-independent \cite{Agnese:2014aze, Akerib:2013tjd, Angloher:2014myn, Aprile:2012nq, Kim:2012rza} and spin-dependent scattering~\cite{Aprile:2013doa, Archambault:2012pm, Behnke:2012ys, Felizardo:2011uw, Kim:2012rza}.
%(LUX~\cite{Akerib:2013tjd}, XENON100~\cite{ Aprile:2011hi}, XENON10~\cite{ Angle:2011th}, CDMS-II(Ge)~\cite{ Ahmed:2010wy}, EDELWEISS~\cite{ Armengaud:2011cy}, and KIMS~\cite{ Kim:2012rza}) and spin-dependent cases (COUPP~\cite{ Behnke:2010xt}, ZEPLIN~\cite{ Lebedenko:2009xe}, PICASSO~\cite{ BarnabeHeider:2005pg}, and KIMS~\cite{ Kim:2012rza}). 
New dark matter candidates~\cite{Belikov:2010yi, Draper:2010ew, Buckley:2010ve, Gresham:2013mua}, instrumentation effects~\cite{Hooper:2010uy, Nygren:2011xu}, backgrounds ~\cite{Blum:2011jf, Davis:2014cja, Ralston:2010bd} and modifications on the distributions of the local dark matter halo~\cite{Fox:2010bz, Lee:2013wza} have all been proposed in an attempt to reconcile these seemingly contradictory results with limited success. %No single source of background can conclusively explain the observed modulation~\cite{ Bernabei:2013xsa}.

A large radio-pure array of NaI(Tl) crystals placed deep in the Antarctic ice near the South Pole will have the ability to directly test DAMA's claim~\cite{Cherwinka:2011ij}. The expected dark matter modulation has a constant phase everywhere on Earth, whereas any modulation resulting from seasonal effects reverses its phase between the Northern and Southern Hemispheres.  The South Pole ice offers up to 2800\,m of overburden, is extremely radio-pure, and strongly suppresses the effects of environmental and seasonal variations such as temperature, humidity, and pressure. Furthermore, the Amundson-Scott South Pole Station and IceCube Neutrino Detector provide technical infrastructure and muon coincidence capabilities~\cite{Achterberg:2006md}.

%% Lauren suggested removal of this.
%To mitigate detector effects (nuclear recoil energy and quenching for example), there are several groups testing DAMA's result by using the same target material, thallium-doped sodium iodide, NaI(Tl). The large signal observed by DAMA implies that a detector of similar size and background will see the modulation to high significance within an exposure of 500\kgyr\ \cite{ Cherwinka:2011ij}. ANAIS is currently under construction at LSC (Laboratorio Subterr\'{a}neo de Canfranc) \cite{ Amare:2006ri, Amare:2012ex, Cebrian201260, Cuesta:2013vpa, Amare:2013lca} with a similar target mass as DAMA. The KIMS experiment~\cite{ Kim:2012rza}, located in the Yangyang Underground Laboratory in South Korea, is testing the hypothesis that the signal observed in DAMA is dominated by interactions with iodine by using roughly 100\,kg of CsI(Tl).
%
% !TEX root = dm-ice17_performance.tex
% Just give the technical details on the hardware

\section{Experimental Setup}
\label{sec:dmice}

%%%%%%%%%%%%%%%%%%%%%%%%%%%%%%%%%%%%%%%%%%%%%%%%%%%
\subsection{Detector}
\label{sec:dmice:detector}
\begin{figure}[!b]
	\includegraphics[height=0.21\textheight]{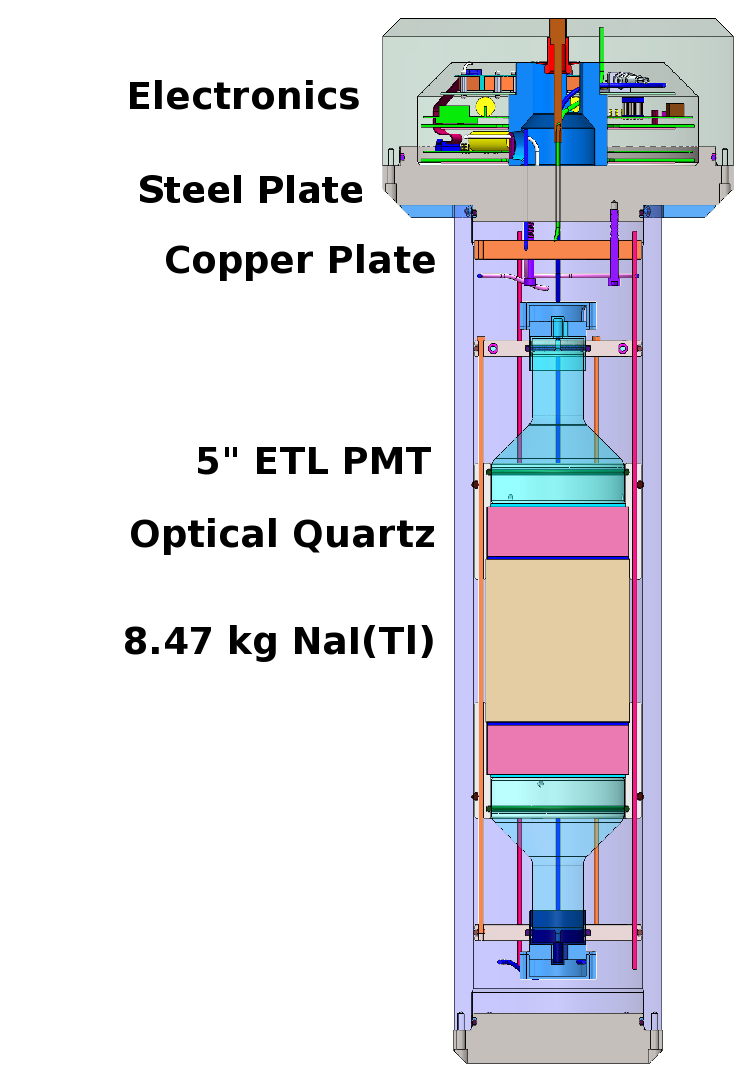}
	\includegraphics[height=0.21\textheight]{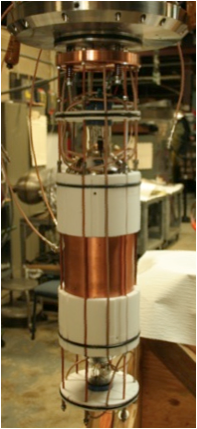}
	\includegraphics[height=0.21\textheight]{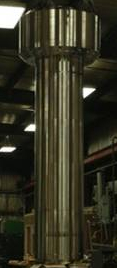}
	\caption{Engineering drawing (left) and photographs of one of the \dm\ detectors with (right) and without (center) the stainless steel pressure vessel. The wider upper section of the stainless steel vessel contains the digitizing and control electronics, high voltage generation board, and isolation transformer for each PMT. The lower section houses one crystal (light brown in left fig) coupled to two PMTs (light blue) via quartz light guides (pink). Power and communication to the detector is provided through a ``special devices" breakout from the main IceCube communication cable. The connection to the detector is made through a single leak-tight penetrator at the top of the pressure vessel.}
	\label{fig:detector}
\end{figure}

\dm\ consists of two 8.47\kg\ NaI(Tl) scintillating crystals; each crystal is optically coupled to two photomultiplier tubes (PMTs) through quartz light guides (see \fig{fig:detector}). Each detector assembly (denoted \dma\ and \dmb), along with its data acquisition and control electronics (see~\sect{sec:dmice:elec}), is housed in its own stainless steel pressure vessel and encased in the South Pole ice at a depth of 2457\,m (see~\sect{sec:dmice:location}). The NaI(Tl) crystals, light guides, and photomultiplier tubes are those used by the former NaIAD experiment that ran from 2000\,--\,2003 at the Boulby Underground Laboratory~\cite{ Ahmed:2003su, Alner:2005kt}.  They were stored in sealed copper boxes at the Boulby Underground Laboratory from 2003\,--\,2010, when they were retrieved to be used in \dm.

The two NaI(Tl) crystals, denoted DM80 (\dma) and DM81 (\dmb) in NaIAD publications, were produced by Bicron and encapsulated by Saint-Gobain. Previously measured background rates at 6\,--\,7\kevee\ of 7\,--\,10\dru~\cite{ Ahmed:2003su, Alner:2005kt} are consistent with those observed in \dm\ and reported in this paper (see~\sect{sec:bkgd}). These crystals provide well characterized detectors for this experiment; however, the high background rates limit the sensitivity of \dm.

Each cylindrical crystal measures 14.0\,cm in diameter and 15.0\,cm in length. For diffusive light reflection, the crystals are wrapped in thin sheets of polytetrafluoroethylene (PTFE).  They are encapsulated with an oxygen-free high thermal conductivity (OFHC) copper housing and a thin quartz window on either end to protect the hygroscopic crystal. The encapsulation has an outer diameter of 14.6\,cm and length of 16.5\,cm.

The scintillation light is recorded by 5-inch 9390-UKB PMTs with C636-KFP voltage divider bases (Electron Tubes Limited).  The crystals are shielded from the radioactivity in the PMTs with 5.0\,cm thick quartz light guides. A 3\,mm layer of Q900 silicone gel (Quantum Silicones) coated in EJ-550 optical grease (Eljen Technology) provides optical coupling between the interfaces of the PMTs, light guides, and NaI(Tl) crystals.  The gel also helped suppress mechanical shock between components during shipping and deployment. Q900, used in IceCube to couple the PMTs to the pressure housings, has been demonstrated to be robust, clean, and stable for long-term operation at the South Pole.

PTFE tubes (Applied Plastics Technology) around the crystal, light guides, and PMTs serve to reflect light and provide mechanical support against lateral movements.  A PTFE disk around the base of each PMT maintains the alignment while Buna-N O-rings on the PTFE tubes and disks hold the detector centered in the pressure vessel. The optical components are both compressed together and protected from mechanical shocks by a series of springs supported by six threaded OFHC copper rods.

The detector assembly is suspended from an OFHC copper plate (orange in~\fig{fig:detector} left) which in turn hangs from a stainless steel midplate (gray in~\fig{fig:detector} left). Holes for the wires were drilled at $45^{\circ}$ through the steel midplate to avoid direct line of sight to the optical components. The lower pressure vessel volume was flushed with dry nitrogen and the holes were potted with silicone-based epoxy; this sealed off the sensitive detector from the upper pressure vessel volume housing the electronics boards (see~\sect{sec:dmice:elec}). A stainless steel cylinder (dark blue in~\fig{fig:detector}) supports the electronics boards and adds mechanical strength to the top surface of the pressure vessel by transferring some of the load onto the midplate.

Each detector is housed in a stainless steel pressure vessel designed to withstand 10000\,psi of external pressure.  Pressure spikes exceeding 7000\,psi have been observed by IceCube as the drill water column freezes (see~\sect{sec:dmice:location}). SANMAC SAF 2205 stainless steel tube stock (Sandvik) was used for the main body of the pressure vessel. The vendor was chosen as it is known to produce clean stainless steel, and the alloy was chosen for its mechanical properties (yield strength 0.2\per\,=\,450\,MPa, tensile strength\,=\,680\,MPa). The seals were modeled after those used in IceCube's drill head and were hydrostatically tested to 7000\,psi at the University of Wisconsin\,--\,Madison Physical Sciences Laboratory (PSL).

\begin{figure*}[!ht]
	\includegraphics[width=0.95\textwidth]{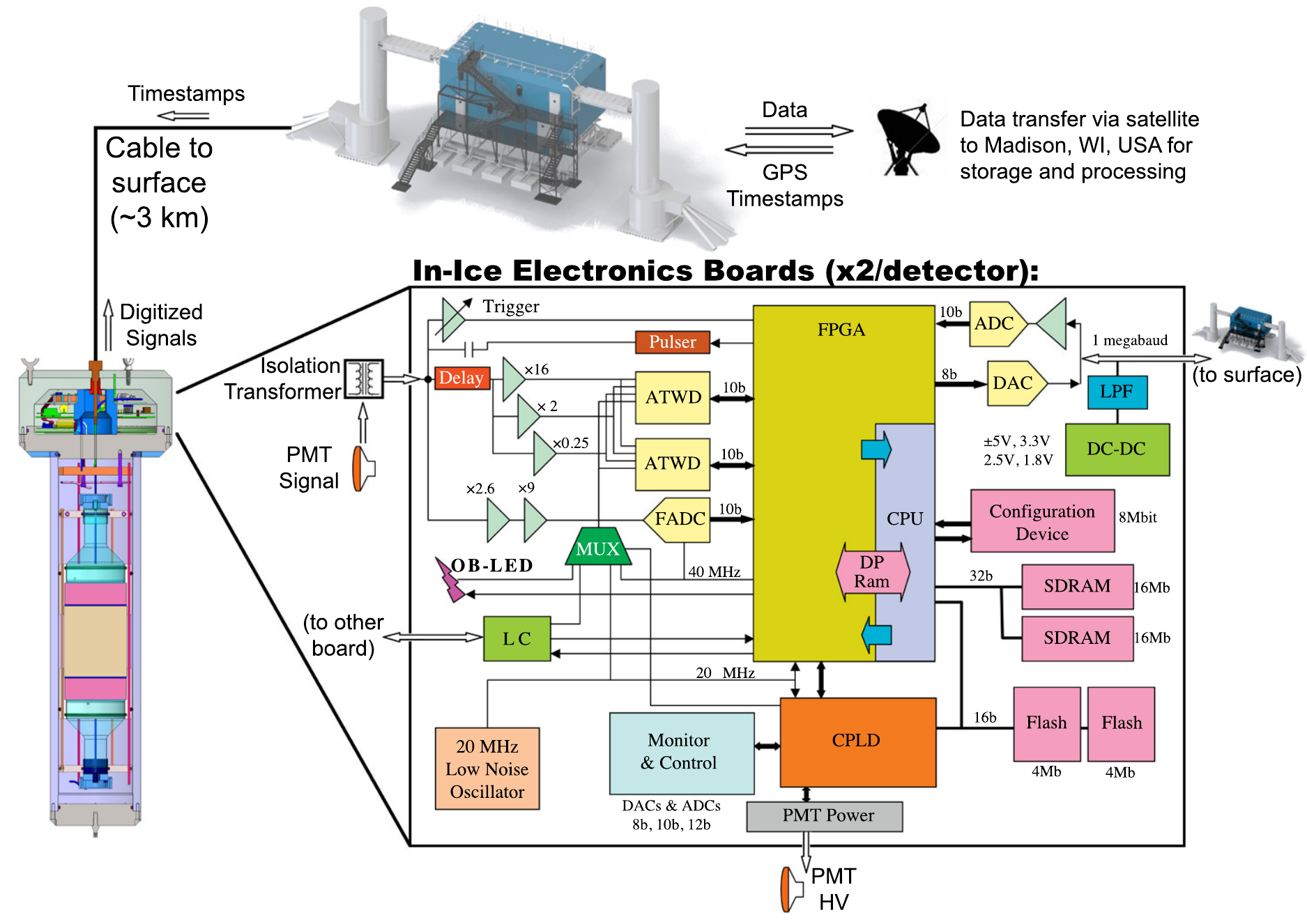}
	\caption{Block diagram of one \dm\ detector and data transfer. The scintillation light is collected with two PMTs and read out by the two IceCube mainboards (DOM-MBs, described in Ref.~\cite{ Abbasi:2008aa}) located in the top portion of the pressure vessel. The waveforms are digitized, given a timestamp, and sent to a hub in the ICL through a twisted pair of copper wires. Data are sent via TDRSS satellite to the data warehouse located at the University of Wisconsin--Madison. The DOM-MB is also responsible for controlling the trigger threshold and PMT HV, as well as monitoring relevant PMT and environmental parameters.} 
	\label{fig:daq} 
\end{figure*}

Low-background counting was performed at SNOLAB~\cite{ Lawson:2011zz}, with results summarized in~\tab{tab:bkgd:sno}.  This counting included drill water, silicone gel, and excess stock from copper, stainless steel, and PTFE. It was not possible to screen materials prior to assembly due to the severe time constraints imposed by IceCube's deployment schedule; therefore, detector components and materials were selected from vendors known to produce radio-clean products. All machined components (e.g. pressure vessel, copper rods, screws) were cleaned using ultrahigh vacuum (UHV) cleaning techniques; the optical components were cleaned with methanol and deionized water. The assembly of detector components was performed in the semiclean room at PSL that was used for assembling IceCube modules.

%%%%%%%%%%%%%%%%%%%%%%%%%%%%%%%%%%%%%%%%%%%%%%%%%%%
\subsection{Electronics and DAQ}
\label{sec:dmice:elec}

Each PMT is independently controlled and monitored by its own IceCube digital optical module mainboard (DOM-MB) assembly and high voltage (HV) board~\cite{ Abbasi:2008aa} located in the top portion of the pressure vessel (see~\fig{fig:daq}).  The DOM-MBs are controlled remotely through a hub located in the IceCube Laboratory (ICL) on the surface of the ice. The hubs are remotely accessible via TCP/IP when there is a satellite connection to the South Pole Station or by the on-site personnel at any time.

The signal from each PMT is carried on an RG303 cable to the respective DOM-MB.  The signal cables have the same length to match their signal transit times. The DOM-MBs contain a field programmable gate array (FPGA) that is configured with a simplified version of IceCube's data acquisition (DAQ) software~\cite{ Abbasi:2008aa}.  Analog PMT signals passing the trigger conditions are digitized \textit{in situ} by the DOM-MBs before begin transmitted to the hub. The hub clock is synchronized to the GPS receiver at the ICL and translates the DOM-MB timestamp into universal time, correcting for the cable lengths of 2500 and 2995\,m for \dma\ and \dmb, respectively.

The signal from each PMT is digitized in four separate channels to allow for different time windows and a broad dynamic range. The flash analog to digital converter (FADC) channel collects 255~samples at 40\,MHz over a 6.375\usec\ window. The FADC channel has a gain of $\times$23.4 and 10-bit dynamic range.  The three analog transient waveform digitizer (ATWD) channels each collect up to 128~samples at a programmable sampling rate of 100\,--\,500\,MHz.  Each ATWD channel has a 10-bit dynamic range but different gain: ``ATWD0'' at $\times$16, ``ATWD1'' at  $\times$2, and ``ATWD2'' at $\times$0.25.  Together the ATWD channels offer an effective dynamic range of 14 bits, allowing waveforms to be recorded without saturation from single photoelectrons (sub-keV) up to \gt 20\,MeV.

Although programmable, the run settings were held constant for all physics data presented here. The PMT HV settings were \{1000, 1000, 1100, 950\}\,V for \{PMT-1a, PMT-1b, PMT-2a, PMT-2b\}.  PMT-1a and PMT-1b are the upper and lower PMTs coupled to \dma\ respectively (likewise for PMT-2a and PMT-2b in \dmb). The trigger threshold of each PMT was set to \aprox 0.25 photoelectrons. A coincidence window of \aprox 400\,ns was imposed before recording the waveforms of the two PMTs coupled to the same crystal. The sampling rate of the \dm\ ATWD chips was set at \aprox 210\,MHz for a digitized waveform time window of \aprox 600\,ns. A \aprox 0.7\,ms deadtime is introduced during waveform digitization.

The DAQ records hardware monitoring information including temperature and pressure (measured at the DOM-MB) as well as PMT high voltage and trigger rate.  Monitoring information, collected regularly during data taking, was recorded every 2\s\ from January 2011 to February 2012 and every 60\s\ since then.  The monitored parameters are discussed in~\sect{sec:dds} and the associated noise is discussed in~\sect{sec:bkgd:noise}.

%%%%%%%%%%%%%%%%%%%%%%%%%%%%%%%%%%%%%%%%%%%%%%%%%%%
\subsection{Location}
\label{sec:dmice:location}

\begin{figure}[!ht]
	\includegraphics[width=\hw]{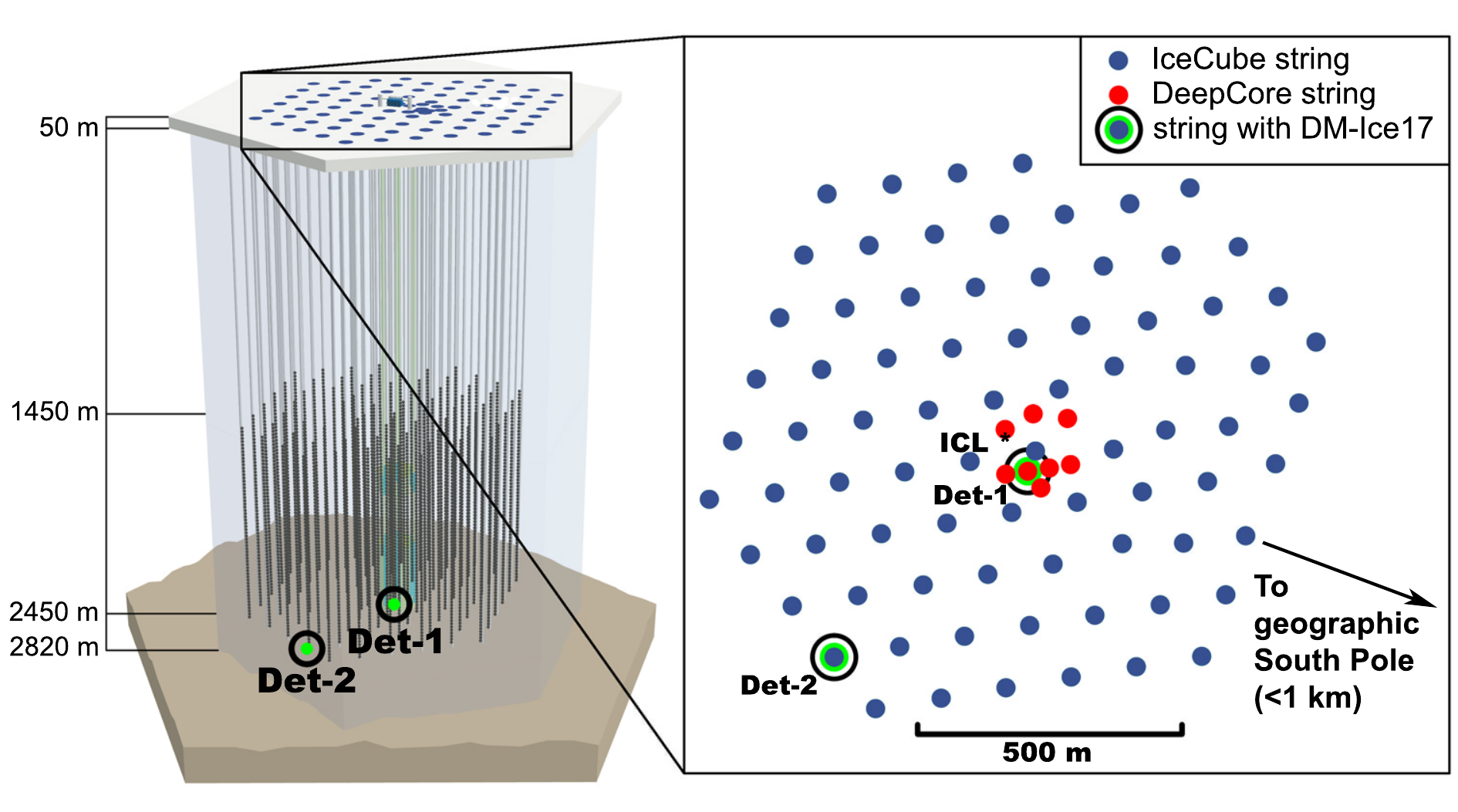}
	\caption{The locations of the \dm\ detectors within the IceCube array, with \dma\ near the array center and \dmb\ at the periphery.  Both detectors are at a depth of 2457\,m and are horizontally separated by 545\,m.  In the array top-view, blue points are IceCube strings, and red points are DeepCore strings.}
	\label{fig:ice-array}
\end{figure}

\dm\ is operating in Antarctica roughly 1\,km from the geographic South Pole (see~\fig{fig:ice-array}). It was codeployed with the final seven strings of IceCube~\cite{ Achterberg:2006md} during the construction season of the 2010/2011 austral summer. The two \dm\ detectors were lowered into separate, water-filled holes, each 2457\,m deep and 60\,cm in diameter, and permanently frozen into place.  These holes, drilled by Ice Cube's enhanced hot water drill, are separated by 545\,m. The ice provides 2200 meters water equivalent of overburden.

Both detectors are deployed on IceCube strings, 7\,m below the lowest digital optical module (DOM). The IceCube array consists of 1\,km$^{3}$ of glacial ice instrumented with 5160 DOMs distributed across 86 strings, with 324 additional DOMs on the surface of the ice as IceTop. One of the DM-Ice17 detectors, \dma, is located near the center of the IceCube array at the bottom of DeepCore~\cite{DeepCore:2011ym} on String-79~\cite{IceCube:2013map}. \dmb\ is at the edge of the IceCube array on String-07. Though operating independently from IceCube, \dm\ uses the same DAQ software and GPS time stamps, providing opportunities to study coincident events with IceCube. The power and communication to the \dm\ detectors are established through twisted pairs of copper wires connected to IceCube's main signal cables.  They are controlled via hubs in the ICL located on the surface of the ice.

The South Pole ice is 90,000\,--\,100,000 years old at depths of 2400\,--\,2500\,m~\cite{Ackermann:2013mbl}. This highly pure glacial ice contains \aprox 10\,ppb of dust, mostly consisting of volcanic ash; the total contamination level from these impurities is \aprox 10$^{-5}$\,ppb for \iso{U}{238} and \iso{Th}{232} and \aprox 10$^{-2}$\,ppb for \iso{K}{nat}~\cite{Rose:1981}. Optical scattering measurements made by IceCube precisely quantify the dust concentrations throughout their detector array, allowing the correlation of South Pole ice with Antarctic ice cores. The \iso{U}{238} concentration was validated by the Vostok ice cores using inductively coupled plasma sector field mass spectrometry (ICP-SFMS)~\cite{Gabrielli:2005}.

The bedrock's contribution to the background of the detectors is negligible as they are shielded by over 300\,m of ice. The environmental radon is also expected to be a negligible component of the total background as the detectors are completely encased in ice. The contamination levels of the water that fills the drilled holes were measured with HPGe counting. Its simulated contribution to the region of interest for dark matter searches was found to be negligible (see~\sect{sec:bkgd:bkgd}).
%
% !TEX root = dm-ice17_performance.tex
% Talk about the external/environmental aspects of the detector

\section{Detector Deployment and Performance}
\label{sec:dds}

%%%%%%%%%%%%%%%%%%%%%%%%%%%%%%%%%%%%%%%%%%%%%%%%%%%
\subsection{Deployment}
\label{sec:dds:deploy}

The detectors were shipped pre-assembled in November 2010, first by land to Los Angeles, CA, then by air to Christchurch, New Zealand.  The detectors waited in Christchurch for two weeks on the surface before being flown to McMurdo Station in Antarctica. After a day in McMurdo, the detectors were flown to the South Pole Station. Time spent on the surface at the South Pole (2835\,m elevation) was minimized to limit exposure to the higher cosmic ray rates.

The sealed pressure vessels were strapped down vertically in custom-designed wooden shipping crates. A Shock Timer Plus 3D sensor (Instrumented Sensor Technology) was mounted inside each crate to monitor temperature, humidity, and mechanical shock during shipment.  The large mass of the pressure vessel and additional layers of insulation mitigated thermal shock while the suspension system within the pressure vessel dampened mechanical shock to the crystals (see~\sect{sec:dmice:detector}). Bare NaI(Tl) crystals can handle surface temperature gradients of \gt 50\,\deg~\cite{NASA:1984}, but encapsulated detectors are rated to only 8\,\deg/hr due to different coefficients of thermal expansion of the crystals and encapsulation materials.

From Wisconsin to New Zealand, the temperature varied between 4 and 24\,\deg\ with a rate of change less than 3\,\deg/hr. Upon arrival at McMurdo Station, the temperature dropped from 25\,\deg\ to \nobreakdash-7\,\deg\ over 15\,hrs. At the South Pole, the temperature dropped 20\,\deg\ in its first 7\,hrs before stabilizing at the surface temperature of roughly \nobreakdash-25\,\deg.

Calculations of the thermal conduction of the pressure vessel were made prior to deployment to verify that safe thermal gradients would be experienced at all times, especially during the rapid temperature equilibration upon detector entry into the 0\,\deg\ water at deployment. The temperature shock was minimized by allowing the detectors to thermalize to 10\,\deg\ in the deployment tower prior to being lowered rapidly (\lt 30\,min) through the \nobreakdash-50\,\deg\ air column in the top \aprox 50\,m of the hole.

Thermal variations measured during shipment and storage were \lt 6\,\deg/hr prior to detector deployment. The largest mechanical shocks recorded were \lt 10\gs\ for both \dma\ and \dmb\ and were primarily during commercial shipment to New Zealand.  No damage to the detectors or loss of light collection efficiency can be discerned from the performance of the detectors.

%%%%%%%%%%%%%%%%%%%%%%%%%%%%%%%%%%%%%%%%%%%%%%%%%%%
\subsection{Freeze-in and temperature stability}
\label{sec:dds:temp}

\begin{figure}[!ht]
	\includegraphics[width=\hw]{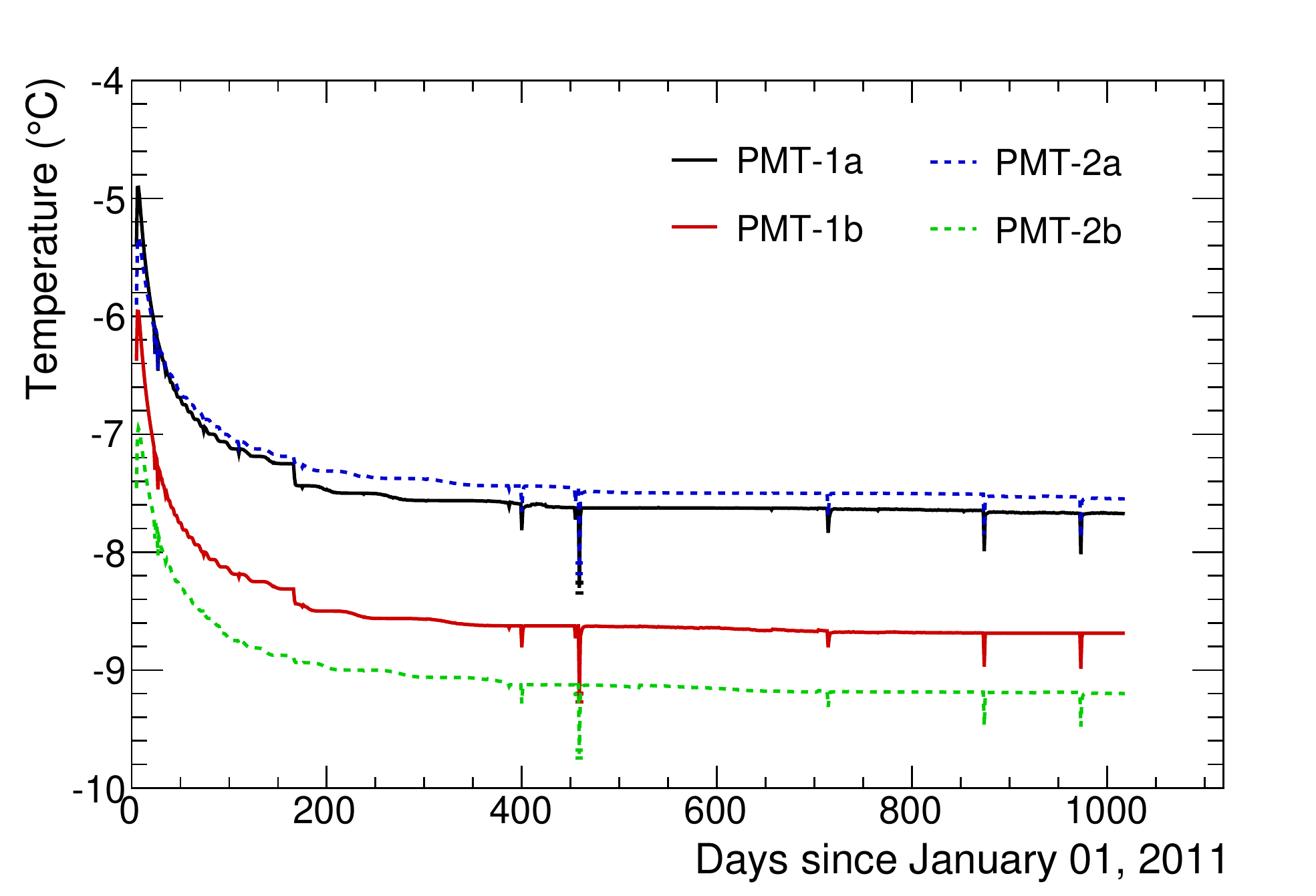}
	\caption{Temperature recorded by \dm\ mainboards since January 01, 2011, sampled every 2 sec through February 2012 and every 60 sec after (see~\sect{sec:dmice:elec}). Two time constants are observed: a relative fast component (\aprox 10\,days) that relates to the freeze-in process and a slower time constant (\aprox 100\,days) as the heat accumulated in the ice surrounding the hole during drilling is dissipated. Discontinuities observed correspond to the PMT high voltage settings change (day~\aprox 160) and subsequent power outages (negative spikes).}
	\label{fig:temps} 
\end{figure}

%The crystals thermalize to the temperature of the surrounding ice at roughly $-18\,\deg$~\cite{Price:2002temp}.
The temperatures of the DOM-MBs are read out as part of the regular monitoring routine (see~\fig{fig:temps}). The recorded temperatures are \aprox 10\,\deg\ warmer than the surrounding \nobreakdash-18\,\deg\ ice due to the \aprox 3.5\,W dissipated by the electronics~\cite{ Abbasi:2008aa}.  The two DOM-MBs in the same pressure vessel are stacked on top of each other and rotated by 60$^{\circ}$ with respect to one another. The temperature sensor on the top DOM-MB sees a 2\,--\,3\,\deg\ higher temperature than its partner from the dissipated heat of the DC-DC converter located directly below it on the lower DOM-MB.

As the ice froze and the detectors thermalized, the temperature decrease exhibited a fast and a slow exponential time component. The fast freeze-in time constant (\aprox 10\,days) is due to the water in the drilled hole freezing around the detector. The slow component of thermalization (\aprox 100\,days) is thought to be due to the thermalization of the heat deposited in the nearby ice during the hot-water drilling process. The small drop on day \aprox160 corresponds to the change in the power dissipation in the mainboard when the PMT high voltage was lowered to the final run settings. The occasional dips in temperature correspond to power outages in either the ICL or the hub affecting the mainboards for more than an hour.

During the two-year physics data run, the temperature of the \dm\ MBs has been stable with a gradual cooling of \aprox 0.25\,\deg\ and an average daily RMS of \aprox 0.02\,\deg.  The thermal mass of surrounding ice provides \dm\ with smaller temperature fluctuations than achieved by similar experiments~\cite{Bernabei:2008yi}.  No temperature modulation is observed within the measurement error. The crystal temperature is expected to be more stable than these DOM-MB measurements due to its larger thermal mass and increased distance from the electronics. The temperature dependence of the detector response is the subject of future studies and is not addressed here.

%%%%%%%%%%%%%%%%%%%%%%%%%%%%%%%%%%%%%%%%%%%%%%%%%%%
\subsection{Hardware stability}
\label{sec:dds:stability}

Regularly recorded PMT high voltages are stable to within fractions of a volt (0.4\,--\,0.8\,V$_{RMS}$) with the exception of PMT-2b. Despite a constant HV set point, PMT-2b shows random variation in monitored HV around the set point (\aprox 30\,V$_{pp}$). No correlations are observed between this HV oscillation and either PMT gain or single photoelectron (SPE) rate.  The monitored HV is stable to within 0.1\,mV (1 in 10$^4$) over the two-year period.

\begin{figure}[!t]
        \includegraphics[width=\hw]{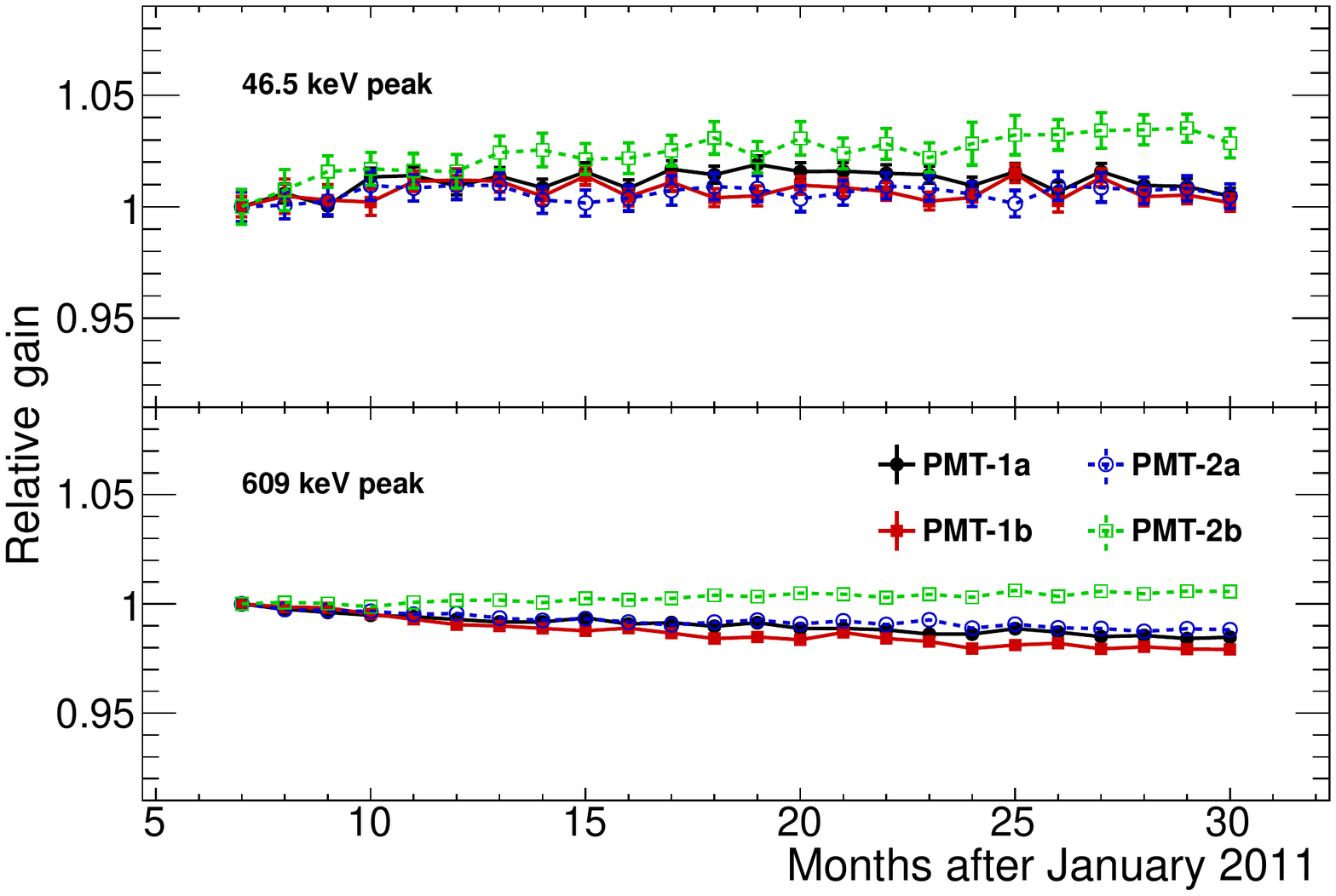}
        \caption{Relative uncalibrated peak position of the 46.5\kev\ \iso{Pb}{210}\ (top) and 609\kev\ \iso{Bi}{214} (bottom) gamma peaks demonstrating \dm\ gain stability. The 46.5\kev\ peak has approximately constant light collection, while the 609\kev\ peak has a \lt 2\per\ decrease over two years.}
        \label{fig:gain}
\end{figure}

The gain stability of the PMTs was quantified from the light collection efficiency at the prominent 609\kev\ and 46.5\kev\ peaks in the uncalibrated energy spectrum (see~\fig{fig:gain}). The 609\kev\ peak from \iso{Bi}{214}\ exhibits a small (\lt 2\per) decrease in light collection over two years. The 46.5\kev\ peak from \iso{Pb}{210}\ exhibits no significant trend with \lt2\per\ fluctuations.  PMT-2b deviates from the trend of the other PMTs at both peaks, showing relative increased gain of a few percent. No time-dependent calibration is applied to the presented data because the gain fluctuations are less than the energy resolution of the detectors; shorter datasets are used for the energy resolution analysis (see~\sect{sec:perf:res}) to limit systematic error of gain drift. No variation is observed in the peak areas within the measurement errors.

%%% Gain stability figure

The total coincident event rate for each detector is \aprox2.5\hz\ and shows a gradual decrease with time (0.027\hz/yr for  \dma\ and 0.013\hz/yr for \dmb).  No variation with time is observed in the PMT trigger thresholds extracted from dark noise (non-coincident) data runs.

%%%%%%%%%%%%%%%%%%%%%%%%%%%%%%%%%%%%%%%%%%%%%%%%%%%
\subsection{Detector livetime}
\label{sec:dds:livetime}

In the two-year dataset presented here, \dm\ took data nearly continuously, with a total livetime of 98.94\per\ (\dma) and 98.92\per\ (\dmb). The downtime is primarily from 10 extended (\gt1 hour) interruptions due to power outages, test runs, and DAQ errors.  Normal uninterrupted data sets achieve a \aprox 99.75\per\ livetime, with small downtime introduced as the DAQ transitions between runs. A \aprox 0.7\,ms deadtime after each event (\aprox2.5\hz\ event rate) introduces an additional \aprox 0.18\per\ downtime.

\section{Detector Response}\label{sec:performance}

%%%%%%%%%%%%%%%%%%%%%%%%%%%%%%%%%%%%%%%%%%%%%%%%%%%
\subsection{Data processing and calibration}
\label{sec:perf:reco}

Events passing the trigger threshold and coincidence window requirements are digitized \textit{in situ} by the DOM-MB (see~\sect{sec:dmice:elec}).  The digitized waveforms are then sent to the hub in the ICL and compiled into one-hour runs.  The data are transmitted via satellite to the University of Wisconsin--Madison for processing.

Offline data processing consists of waveform corrections and energy calibration.  The waveform is first baseline adjusted and then corrected for the frequency response of the passive electronic components.

\begin{figure}[!b]
	\includegraphics[width=\hw]{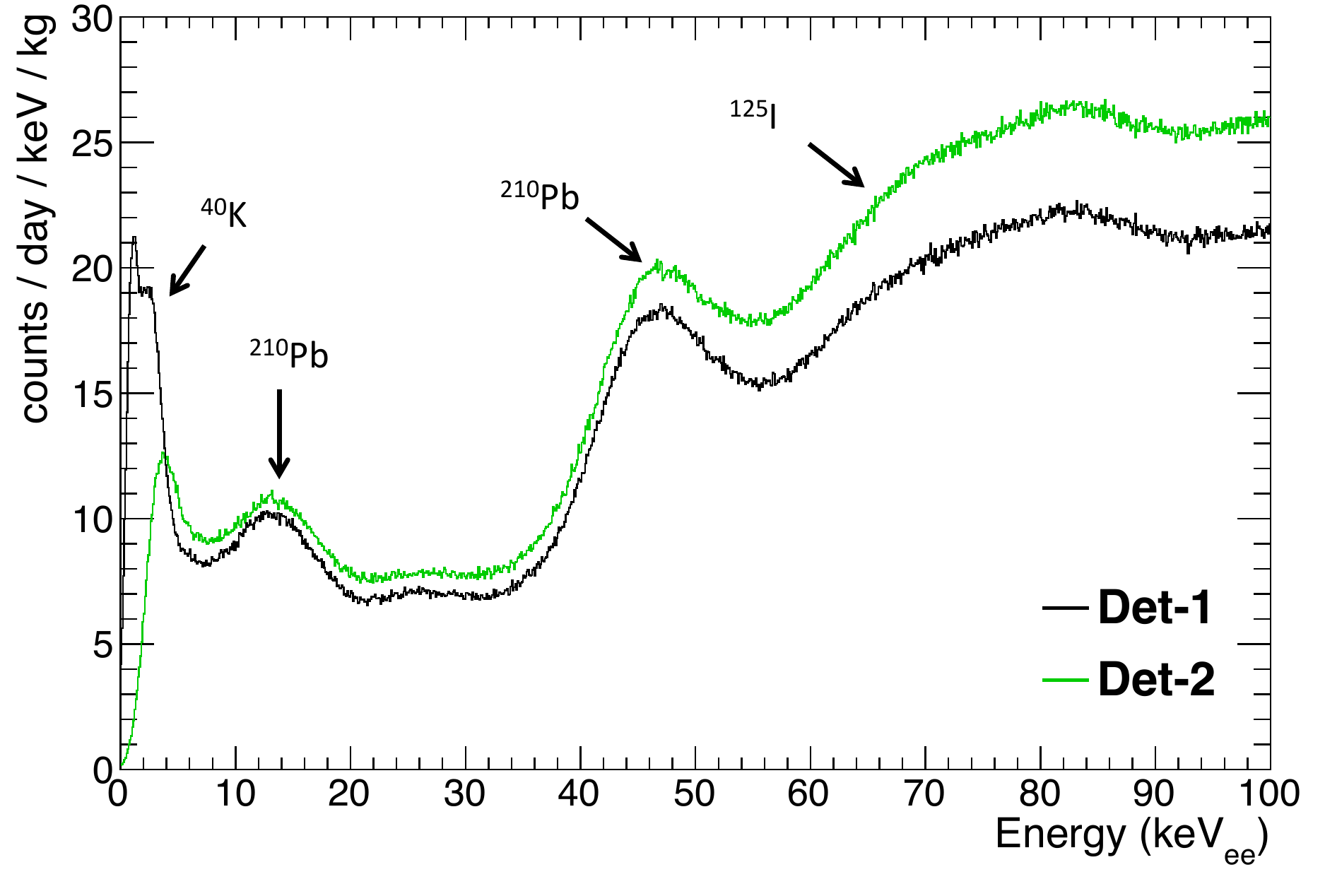}
	\includegraphics[width=\hw]{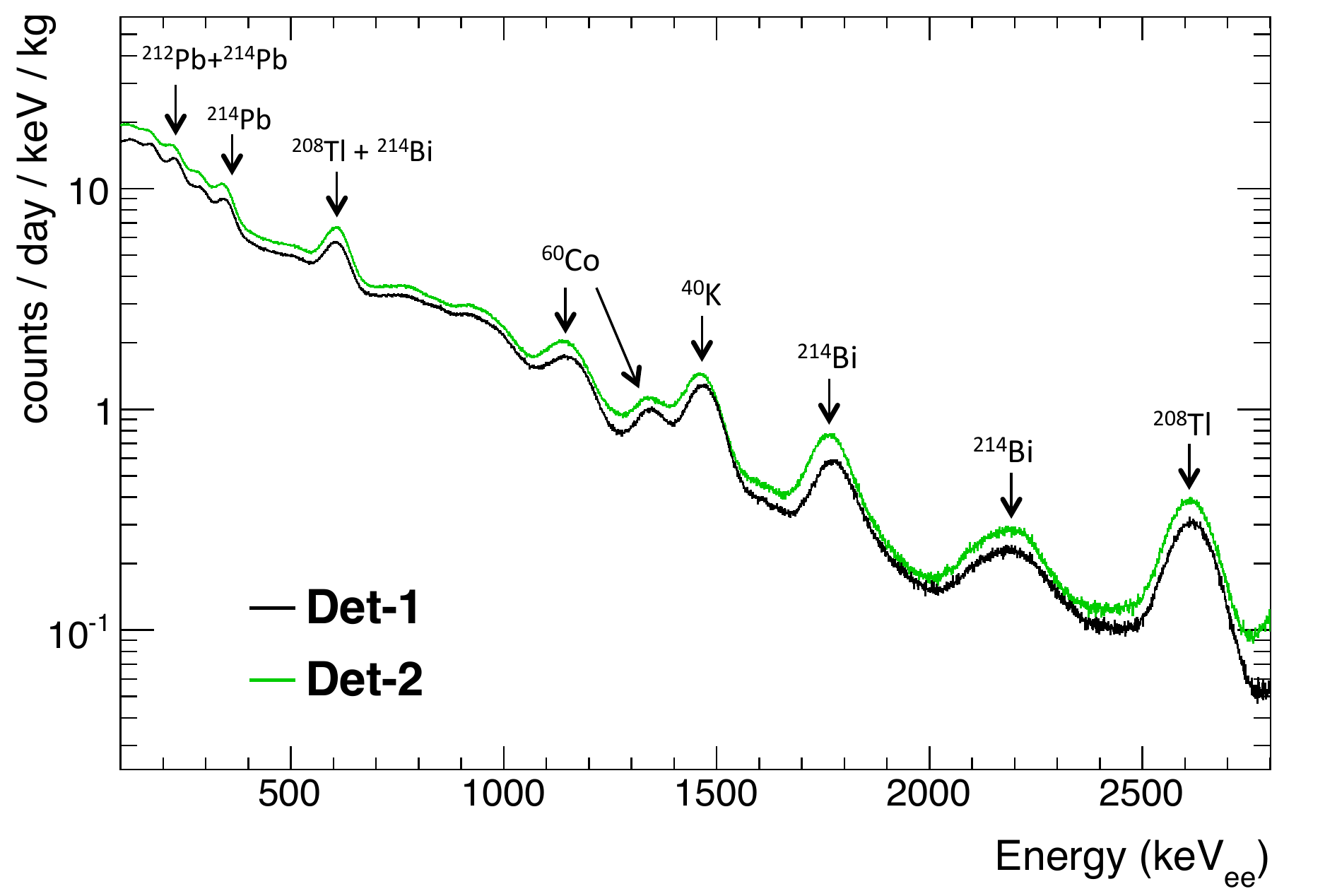}
	\caption{The calibrated energy spectrum with prominent lines identified for both \dma\ (black) and \dmb\ (green). ATWD0 (the highest gain channel) is used for the low-energy plot (top), and ATWD1 is used for the high-energy plot (bottom); the spectra from these two channels overlap well from 100--1000\,\kev. The nonlinear response of NaI requires that separate calibrations be applied for the two energy regions. All lines depicted are included in calibration fits and energy resolution analysis.}
	\label{fig:energy-points-log} 
\end{figure}

The energy calibration begins by integrating the entire corrected waveform (\aprox600\,ns for ATWD).  The resulting spectrum for each PMT is calibrated into \kevee\ using known lines from \iso{K}{40}, \iso{Co}{60}, \iso{I}{125}, \iso{Th}{232}-chain, and \iso{U}{238}-chain.  The final energy spectra (see~\fig{fig:energy-points-log}) are produced by combining the light collected in both PMTs on a single crystal for each event. The energies are reported in electron-equivalent \kev\ (\kevee), unless otherwise stated.

The four digitization channels are optimized for different energy ranges based on their gain and saturation energy.  ATWD0 is useful for 0\,--\,1000\kev, ATWD1 for 100\,--\,5000\kev, and ATWD2 \gt1000\kev. The longer FADC digitization window carries important timing information (see~\sect{sec:bkgd:bipo}), but the FADC energy spectrum is not used in present analyses.

%The FADC spectrum has worse energy resolution due to less refined waveform corrections, but is planned for use in future analyses.

The nonlinear light response of NaI requires that two different energy calibrations be applied for above and below 100 keV (see~\fig{fig:energy-points-log}).  An extrapolation of the high-energy linear calibration results in a negative intercept, consistent with behavior observed in literature~\cite{ Engelkemeir:1956, Khodyuk:2011km}.

%%%%%%%%%%%%%%%%%%%%%%%%%%%%%%%%%%%%%%%%%%%%%%%%%%%
\subsection{Cosmogenic isotopes}
\label{sec:perf:cosmo}

Cosmogenically activated isotopes are produced in the detector components prior to deployment and provide a robust verification of the energy calibration. The activation rates were simulated in ACTIVIA (NaI crystal)~\cite{Back:2007kk} or obtained from literature values (steel pressure vessel)~\cite{LaubensteinAct}. The activation calculation incorporates the duration of the stages of construction, shipment, and deployment, as well as the rate scaling for each geographic location and elevation. The resulting isotopes were propagated through the detector using the \dm\ \textsc{Geant4} simulation package (see~\sect{sec:bkgd:bkgd}). In \dm\ data, \iso{I}{125} from \iso{I}{127} activated in the crystal and \iso{Mn}{54} from \iso{Fe}{56} in the pressure vessel provide the clearest cosmogenic signals due to their energies and half-lives.
 
\begin{figure}[!ht]
	\includegraphics[width=\hw]{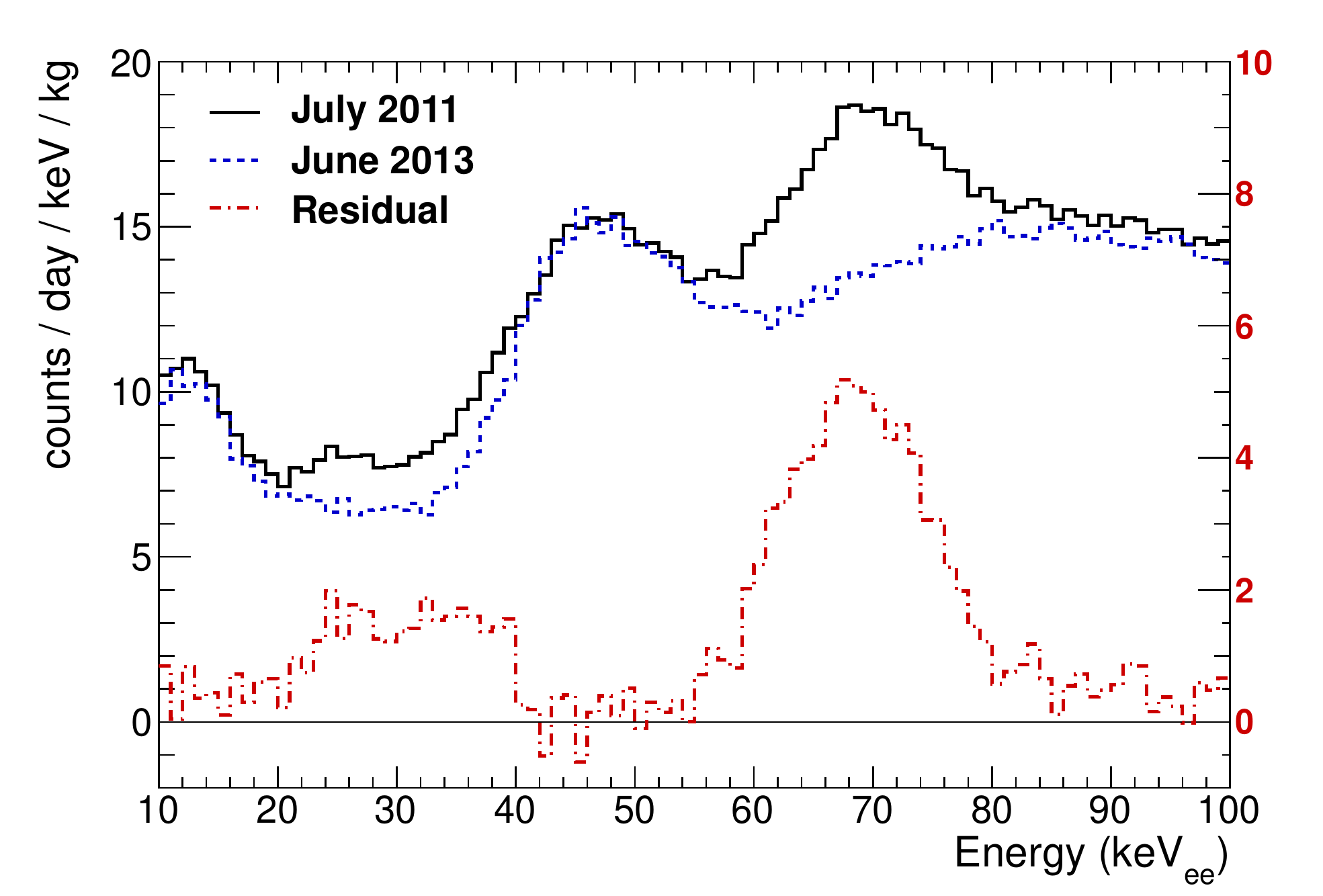}
	\includegraphics[width=\hw]{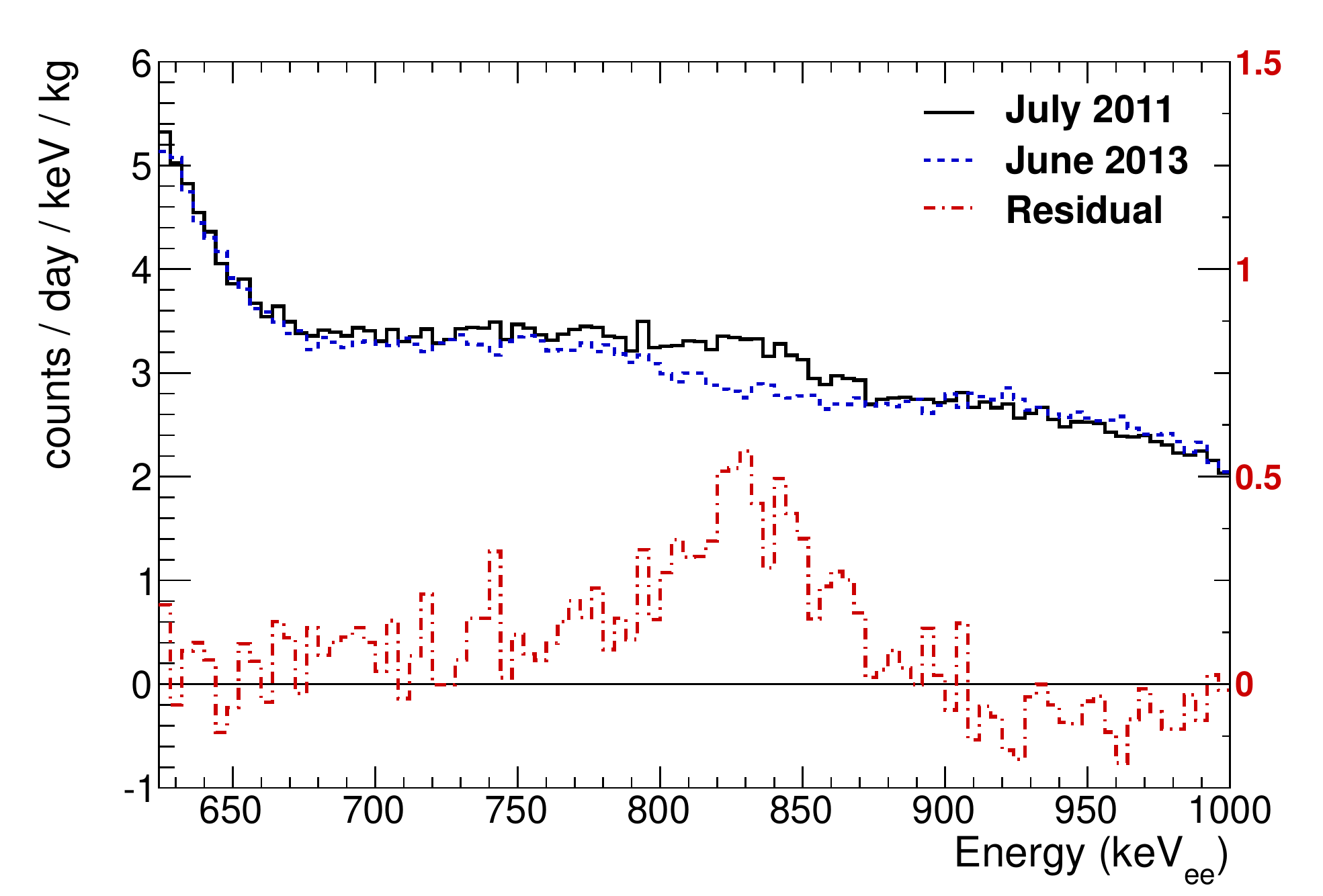}
	\caption{\dma\ spectra from July 2011 (black, solid), June 2013 (blue, dashed), and residual (red, dot-dashed, on rescaled axis) showing evidence for cosmogenic activation.  The top figure shows the decay of the 37.7 and 65.3\kev\ lines from cosmogenic \iso{I}{125}\ in the crystal. The bottom figure shows the decay of the 834.8\,\kev\ gamma line from cosmogenic \iso{Mn}{54} in the pressure vessel.} 
	\label{fig:cosmogenic} 
\end{figure}

Cosmogenic \iso{I}{125} is observed (see~\fig{fig:cosmogenic}), exhibiting the expected peaks at 37.7 and 65.3\kev\ resulting from a combination of gammas and X-rays. Using one-month intervals of data, the measured half-life of this isotope is 59.4\plm 2.7\,days, consistent with the quoted 59.40\plm 0.01\,day half-life of \iso{I}{125}~\cite{nudat}. The \iso{I}{125}\ activity, extrapolated back to deployment, was determined to be 1150\plm 120\,decays/day/kg. Simulation using ACTIVIA estimated an activity of 410\,decays/day/kg. This discrepancy is within the error of the ACTIVIA activation rate~\cite{Back:2007kk} and scaling factor calculations.

The gamma line from  \iso{Mn}{54}\ at 836.1\plm 3.0\kev\ is also observed (see~\fig{fig:cosmogenic}), consistent with the literature value of 834.848\plm 0.003\kev~\cite{nudat}. The measured half-life is consistent with the tabulated value of 312 days, although precise measurement has not been made due to low statistics. The \iso{Mn}{54}\ activity, extrapolated back to deployment assuming the literature half-life, was determined to be 51700\plm 6500\,decays/day (corrected for detection efficiency of decays in the pressure vessel found from \textsc{Geant4} simulation). Simulation based on literature activation rates estimated an activity of 41850\plm 4650\,decays/day (error only from activation rate measurement) from the pressure vessel, in good agreement given additional uncertainty in the scaling factors.

%%%%%%%%%%%%%%%%%%%%%%%%%%%%%%%%%%%%%%%%%%%%%%%%%%%
\subsection{Energy resolution and light yield}
\label{sec:perf:res}

\begin{figure}[!ht]
	\includegraphics[width=\hw]{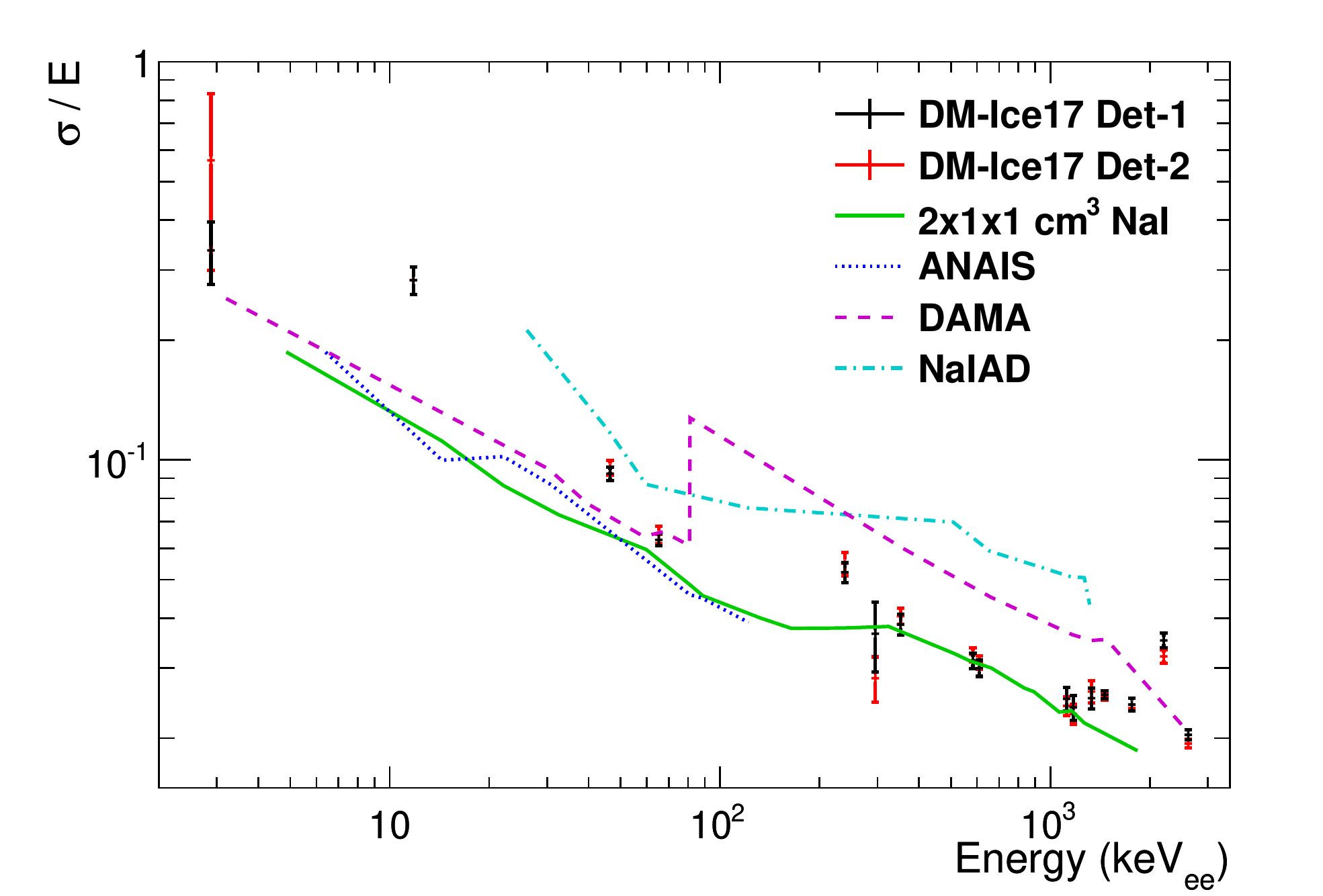}
	\caption{Energy resolution of \dma\ (black) and \dmb\ (red) compared to ANAIS-25 (blue, dotted)~\cite{Amare:2013lca}, DAMA/LIBRA (magenta, dashed)~\cite{Bernabei:2008yh}, NaIAD detector DM77 (cyan, dot-dashed)~\cite{Ahmed:2003su}, and a 2x1x1\,cm$^3$ NaI detector (green, solid)~\cite{sakai:1987}. The discontinuity in the DAMA resolution is due to switching ADC channels at \aprox 80\kev. The internal contamination lines were fit with a Gaussian plus linear background, the quantity $\sigma$ is the fitted standard deviation and the error bar is the fitted parameter error.}
	\label{fig:res}
\end{figure}

The energy resolution of the \dm\ detectors (see~\fig{fig:res}) is measured by fitting a Gaussian plus linear background to the prominent peaks from internal contamination. The resolution measured in the \dm\ detectors is comparable to that of similarly sized NaI detectors from ANAIS-25~\cite{Amare:2013lca}, DAMA/LIBRA~\cite{Bernabei:2008yh}, and NaIAD~\cite{Ahmed:2003su}, as well as to that of a small 2x1x1\,cm$^3$ NaI crystal~\cite{sakai:1987}. The resolution of the 3\kev\ peak from \iso{K}{40} has large error due to the uncertainty in the cut efficiency (see~\sect{sec:bkgd:noise}). The \aprox 14\kev\ peak from \iso{Pb}{210} is broadened due to the many (\gt 10) X-rays that make up that peak. The 2204\kev\ peak from \iso{Bi}{214} is broadened by contributions from \iso{Tl}{208} decay.

Single photoelectron (SPE) events are collected by removing the two-PMT coincidence trigger requirement during regularly scheduled dark noise runs. The light yield is measured by comparing the peak location of the 65.3\kev\ calibration line from \iso{I}{125} to that of the SPE peak. The measured light yields are 5.9\plm 0.1\,pe/keV  and 4.3\plm 0.1\,pe/keV for \dma\ and \dmb\ respectively.

%At low energies, the light yields are 5.9\plm 0.1\,pe/keV (\dma) and 4.3\plm 0.1\,pe/keV (\dmb). The single photo-electron (SPE) spectrum~\cite{ ChirikovZorin:2001qv, Asch:2005pe} was collected during regular dark noise (non-coincident) runs to determine the pe/ADC.  This light yield is then converted into pe/\kev\ for the 65.3\kev\ calibration point from cosmogenic \iso{I}{125}.

%%%%%%%%%%%%%%%%%%%%%%%%%%%%%%%%%%%%%%%%%%%%%%%%%%%
\subsection{Pulse shape discrimination for alphas} \label{sec:perf:psd}

\begin{figure}[!t]
	\includegraphics[width=\hw]{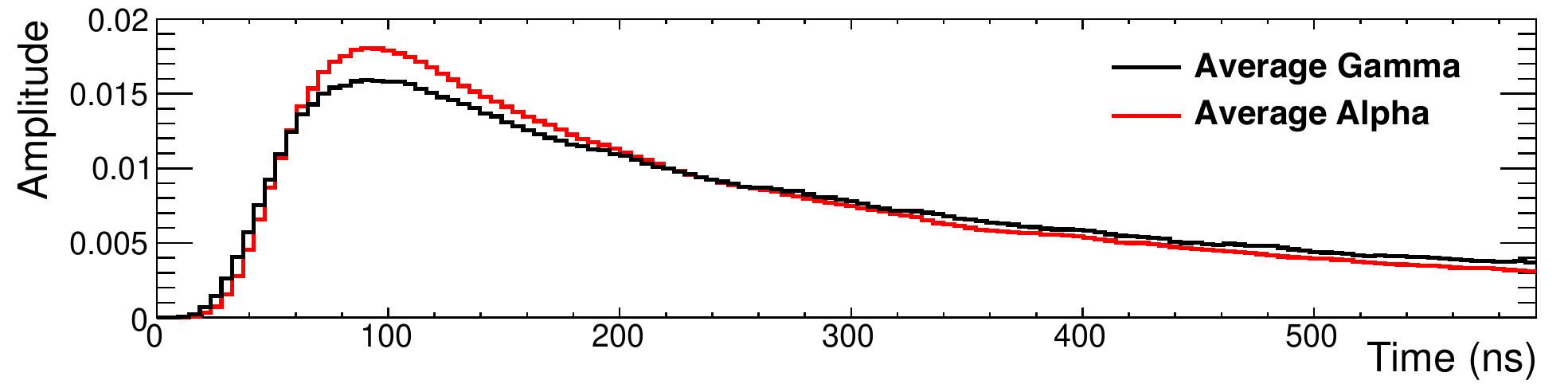}
	\caption{An average of 100 normalized gamma and alpha waveforms as recorded in the ATWD1 channel of PMT-1a. Particle identification is possible through pulse-shape discrimination.}
	\label{fig:bkgd:psd} 
\end{figure}

Alpha events can be separated from gammas via pulse shape discrimination. Alphas produce shorter scintillation waveforms than gammas, leading to observable differences in the behavior in the tail of the recorded waveforms~\cite{Bernabei:2008yh,Cuesta:2013vpa}. A comparison of average alpha and gamma waveforms, as recorded in the ATWD1 channel, is shown in~\fig{fig:bkgd:psd}. Longer time windows are available in the FADC channel; however, the pulses are saturated for MeV events and therefore are not used in this analysis.

\begin{figure}[!ht]
	\includegraphics[width=\hw]{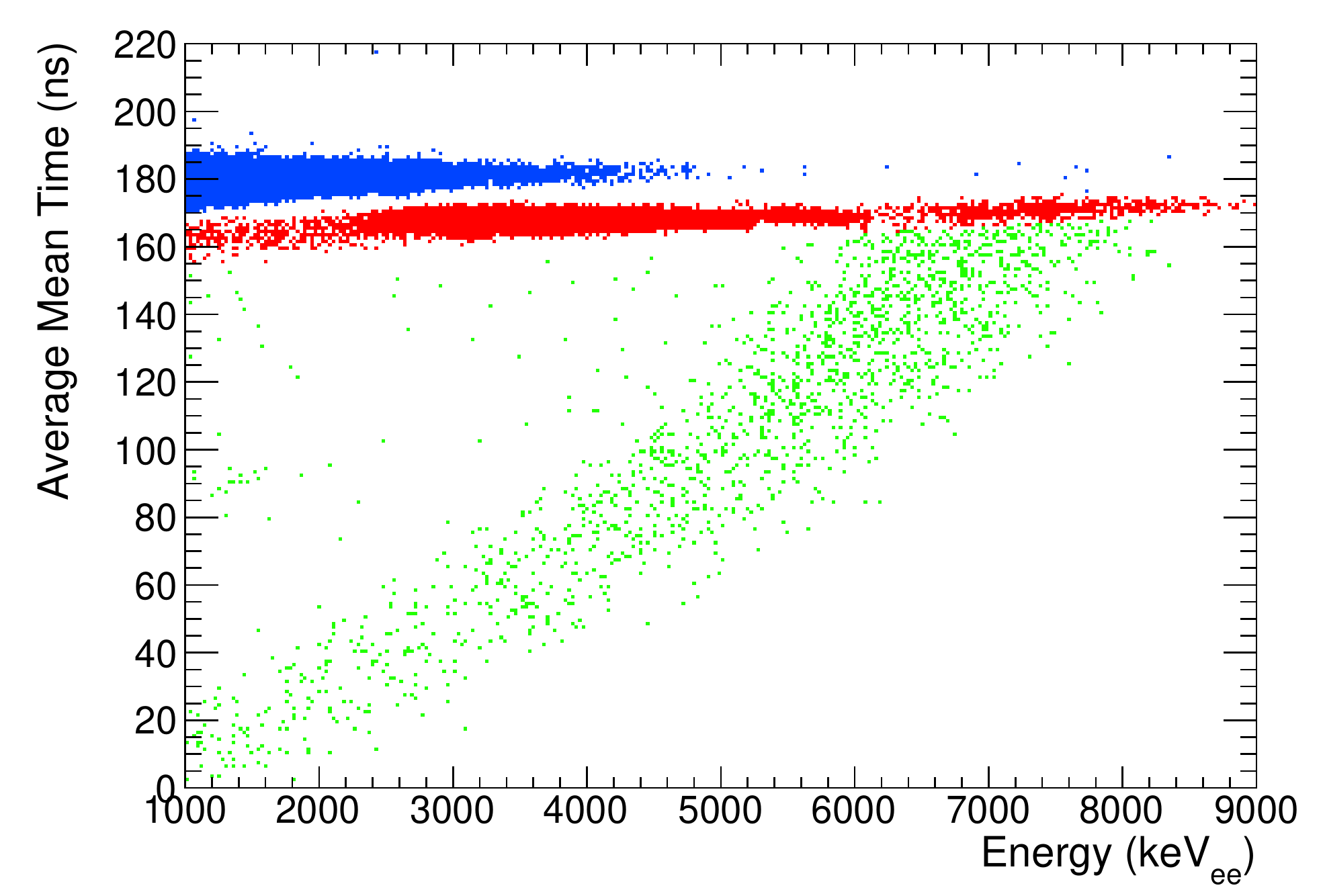}
	\caption{Waveform mean time values from \dma\ demonstrating separation between electron recoils (blue, $\tau \approx 180$\,ns) and alpha events (red, $\tau \approx 165$\,ns).  The diagonal band at lower $\tau$ (green) is composed of Bi-Po events (see~\sect{sec:bkgd:bipo}).}
	\label{fig:tau1} 
\end{figure}

A parameter referred to as the mean time ($\tau$) quantifies the decay behavior of waveforms and is defined numerically as
\begin{equation*}
	\tau = \frac{\sum\limits_{n=n_0}^{n_0+99} \left(n - n_0\right)\left(ADC_n\right)}{\sum\limits_{n=n_0}^{n_0+99} ADC_n},
	\label{equ:meantime}
\end{equation*}
where $ADC_n$ refers to the charge collected in the $n$th time bin of the waveform and $n_0$ is the time when the waveform first reached 50\per\ of its maximum. To account for variations in the time between the start of scintillation and the start of the recorded waveform, the mean time calculation begins at $n_0$ and samples 100 bins.  The separation observed (see~\fig{fig:tau1}) using the mean time parameter is consistent with that reported by other NaI(Tl) experiments~\cite{Bernabei:2008yh, Kim:2014toa}, although our shorter sampling window reduces the calculated mean time.

Above 2000\kevee, gamma and alpha separation via the mean time nears 100\per. These two bands can be accurately represented by Gaussian functions with nonoverlapping tails; in the 2500\,--\,3500\kevee\ region, less than one misidentified event is expected for the 24 months of data presented here.

%
% !TEX root = dm-ice17_performance.tex

\section{Background Analysis} \label{sec:bkgd}
\subsection{External backgrounds} \label{sec:bkgd:ext} \label{sec:bkgd:bkgd}

\begin{figure}[!ht]
	\includegraphics[width=\hw]{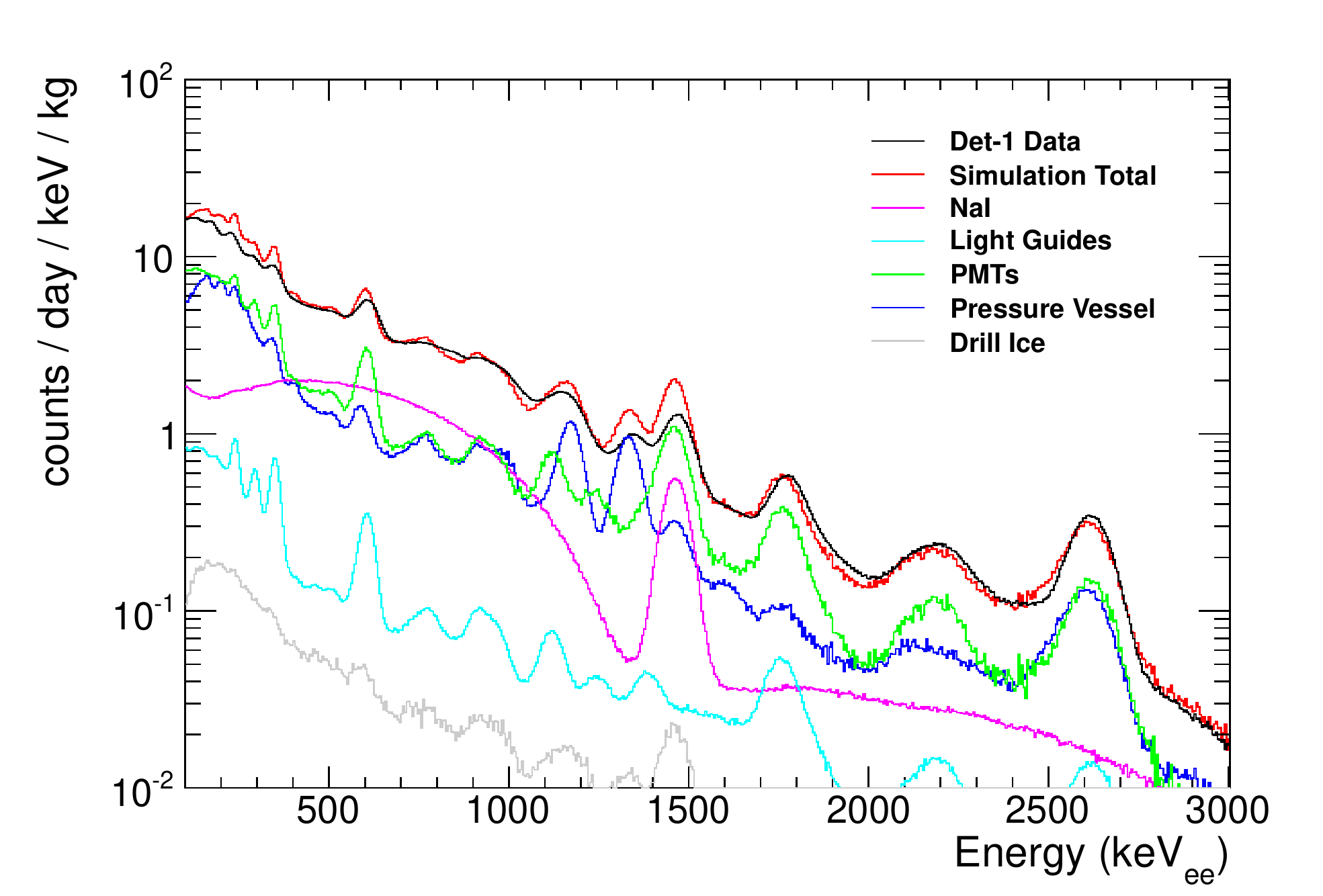}
	\caption{The beta/gamma energy spectrum in \kevee\ from \dma\ (black) and comparison to simulation (red). The simulated contributions of significant detector components are shown separately.  Simulated contaminant levels are derived from data-based estimates and radioassays (see~\tab{tab:bkgd:sno} and~\tab{tab:nai}).} 
	\label{fig:bkgd:gammas} 
\end{figure}

The energy spectrum over 100\,--\,3000\kev\ from the \dm\ data shows good agreement with simulations based on radioassays of components and data-based estimates (see~\fig{fig:bkgd:gammas}). The background levels of the two detectors are comparable, and the spectrum from \dma\ is shown as representative of the two. The contamination levels used in the simulation are listed in~\tab{tab:nai} for the NaI crystals and in~\tab{tab:bkgd:sno} for all other components.

\begin{table*}[!ht]
\begin{ruledtabular}
\caption{Contamination levels of \dm\ detector components in\,\mbk. Quartz measurements are from ILIAS database for `Spectrosil B silica rod'~\cite{ilias}, and PMT levels are from the low-background ETL 9390B datasheet (with increased \iso{U}{238}). Components that were measured specifically for this experiment at SNOLAB~\cite{Lawson:2011zz} are indicated by *.  Components not shown in ~\fig{fig:bkgd:gammas} are indicated by $^\dagger$.}
\label{tab:bkgd:sno}
	\begin{tabular}{l|cccccc}
		Material	& $^{40}$K	& $^{232}$Th	& $^{238}$U	& $^{238}$U	& $^{235}$U	& $^{60}$Co	\\
					&			&			& ($^{234}$Th)	& ($^{226}$Ra)	&			&			\\
		\colrule
		Quartz Light Guides	& 0.50\plm 0.03	& \lt 4.9	& \multicolumn{2}{c}{12}	& ... & ... \\
		{ETL 9390B PMT}	& 9300	& 1000	& \multicolumn{2}{c}{2400}	& ... & ... \\
		Steel Pressure Vessel *	& 13.77\plm 6.38	& 6.49\plm 0.96	& 118.31\plm 60.11	& 2.28\plm 0.72	& 8.79\plm 1.68	& 7.19\plm 0.82	\\
		Drill Ice *		& 3.71\plm 1.36	& 0.55\plm 0.17	& 6.69\plm 3.02		& 0.39\plm 0.14		& 0.38\plm 0.21		&	0.12\plm 0.05	\\
		Silicone Optical Gel * $^\dagger$	& 39.50\plm 18.60	& \lt 0.12	& 2.08\plm 1.10	& 38.50\plm 61.00	& 0.96\plm 1.30	& 0.32\plm 0.42	\\
		PTFE Supports * $^\dagger$	& 0.34\plm 5.09	& 0.52\plm 0.44	& \lt 0.41	& 24.46\plm 21.37	& 1.92\plm 0.72	& \lt 0.089	\\
		Copper Plate * $^\dagger$	& \lt 5.13	& \lt 1.22	& 0.17\plm 0.92	& \lt 0.67	& 3.56\plm 1.79	& \lt 0.12\\
		Glacial Ice $^\dagger$	& \aprox3$\times 10^{-4}$	& \aprox 4$\times10^{-4}$	& \multicolumn{2}{c}{\aprox 10$^{-4}$} & ... & ... \\
	\end{tabular}
\end{ruledtabular}
\end{table*}

During construction, excess stock from the stainless steel pressure vessel, PTFE supports, and copper plate, as well as the silicone gel used for optical coupling, were set aside to be measured at SNOLAB's low-background counting facility~\cite{Lawson:2011zz}. Samples of the drill water were taken just prior to heating and recirculation.  Eleven water samples were taken at varying depths on eight separate holes. The contamination levels measured in the extracted drill water show no significant dependence on the depth or the order in which the holes were drilled; \tab{tab:bkgd:sno} shows the average measured contamination.

The contamination levels for the quartz light guides were taken from the `Spectrosil B silica rod' values on the ILIAS database~\cite{ilias}. The contamination levels for the low-background ETL 9390 PMTs were provided in the vendor data sheet; the \iso{U}{238}-chain level was increased by 95\per\ to better match data. The contamination levels of the NaI crystals were derived from spectral features in data validated by simulation (see~\sect{sec:bkgd:nai}). The PTFE light reflector and copper enclosing the crystal (Saint-Gobain encapsulation) were simulated with bulk contamination levels matching the newer materials (PTFE supports and OFHC copper plate measured at SNOLAB); a surface contamination (0\,--\,10\,$\upmu$m) in the copper of 40\,mBq of \iso{U}{238}-chain was included to better match data (see~\sect{sec:bkgd:lowE}).  The glacial ice contamination is estimated using using dust impurity levels measured by IceCube~\cite{Ackermann:2013mbl} scaled to appropriate contamination levels~\cite{Rose:1981} (see~\sect{sec:dmice:location} and~\cite{Cherwinka:2011ij}). The contributions from contamination in the PTFE supports, copper rods and plate, silicone optical gel, quartz light guides, drill ice, and glacial ice were simulated and found to be insignificant.

Detector and background modeling were performed using the 4.9.5 release of the \textsc{Geant4} software package~\cite{geant4, geant4b}. The physics list utilizes the standard electromagnetic interaction models with atomic relaxation~\cite{geant4c} as demonstrated in the ``rdecay02'' example code.

%After fixing the NaI crystal contaminants (see~\sect{sec:bkgd:nai}), the simulation generally agrees with the prominent gamma lines in data when using the estimated concentrations. Future improvements will include a likelihood fit of the simulation to data. 

%the remaining background levels (see~\tab{tab:bkgd:dmice}) were determined from the strong gamma lines (see~\fig{fig:bkgd:gammas}) in the \dm\ energy spectral data.

%\input{bkgs/dmice17}

%%%%%%%%%%%%%%%%%%%%%%%%%%%%%%%%%%%%%%%%%%%%%%%%%%%
%\dm\ data is compared to simulation to verify our understanding of the energy spectrum and prominent gamma lines. Simulations also provide a consistent background model of the experimental apparatus and verification of the material background measurements.

%To confirm our understanding of the energy spectrum and intrinsic backgrounds, we compare the low energy portion of the observed spectrum with simulation.   The data were fit with a detailed simulation that models the radiogenic contamination on the main componenents of the setup and the surrounding ice.   Before fitting to the model, it was necessary to apply pulse selection criteria to eliminate events below 10 keV which do not arise from physical particle interactions in the crystal.  

%%%%%%%%%%%%%%%%%%%%%%%%%%%%%%%%%%%%%%%%%%%%%%%%%%%
\subsection{Backgrounds from NaI crystals} \label{sec:bkgd:nai}

\begin{figure}[!ht]
	\includegraphics[width=\hw]{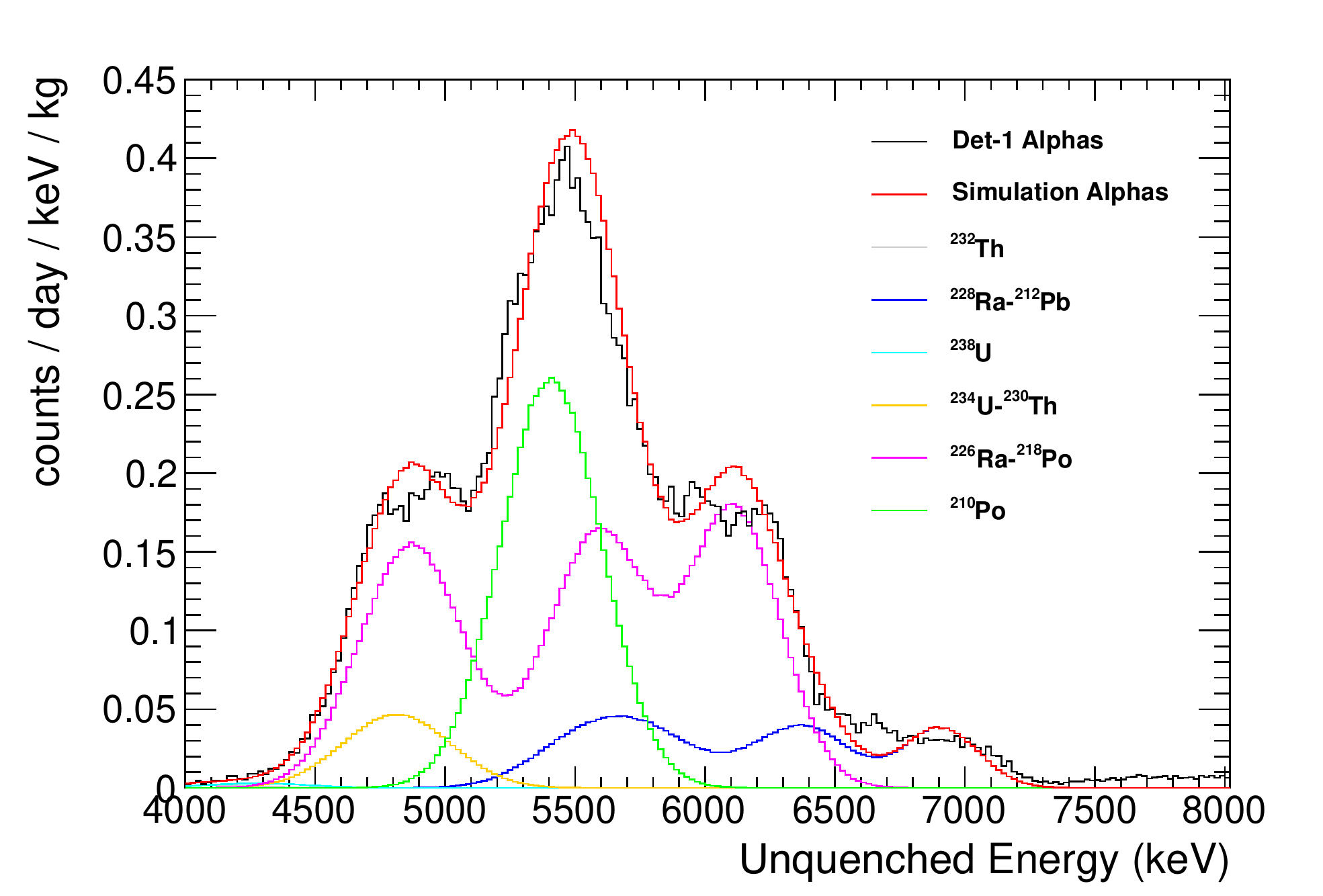}
	\caption{The energy spectrum in the alpha region of \iso{U}{238} and \iso{Th}{232} and chains in the NaI(Tl) crystal as measured with PMT-1b of \dma\ (black) and simulation (red). Comparison to simulation yields contaminant level estimates of the \iso{U}{238}- and \iso{Th}{232}-chains in the crystal (see~\tab{tab:nai}). The energy scale shown takes into account the correction from the gamma-calibrated spectrum using the quenching factor (see text).} %The cause of the shoulders observed in peaks in the alpha spectrum is still under investigation.} 
	\label{fig:bkgd:alphas} 
\end{figure}

The alpha lines between 4.0 and 7.5\,MeV were used to measure the \iso{U}{238} and \iso{Th}{232} contaminations in the NaI(Tl) crystals. The events in these lines are expected to be dominated by those occurring in the bulk of the NaI(Tl) crystals. The spectrum from each PMT is compared to the \textsc{Geant4}-based simulation and contamination levels are evaluated from the respective peak areas (see~\fig{fig:bkgd:alphas}). Both \iso{U}{238}- and \iso{Th}{232}-chains appear to be broken, as described in~\tab{tab:nai}. The cause of the shoulders observed in the peaks of the alpha spectrum is under investigation.
%The alpha events are isolated from gammas primarily using their energy and pulse shape discrimination (see~\sect{sec:bkgd:psd}),

\begin{table}[!ht]
\begin{ruledtabular}
\caption{Contamination in the \dm\ NaI crystals as determined by simulation comparison to data spectral features. The activity levels in the two detectors are consistent within the \aprox 30\per\ error of these numbers. Both $^{238}$U- and $^{232}$Th-chains were observed to be broken.}
\label{tab:nai}
	\begin{tabular}{crcc}
	Isotope	& Subchain		& Activity	\\
			&				& (\mbk)	\\
	\colrule
	\colrule
	$^{40}$K		&			& 17	\\
	\colrule
	$^{129}$I		&			& 1	\\
	\colrule
	\multirow{2}{*}{$^{232}$Th}	& $^{232}$Th	& 0.01	\\
	& $^{228}$Ra\,--\,$^{208}$Tl	& 0.16	\\
	\colrule
	& $^{238}$U\,--\,$^{234}$Pa	& 0.017	\\
	\multirow{2}{*}{$^{238}$U}	& $^{234}$U\,--\,$^{230}$Th	& 0.14	\\
		& $^{226}$Ra\,--\,$^{214}$Po	& 0.90	\\
	& $^{210}$Pb\,--\,$^{210}$Po	& 1.5	\\
	\end{tabular}
\end{ruledtabular}
\end{table}

The scintillation yield of the NaI(Tl) detectors is lower for alpha than for gamma interactions. The measured alpha quenching factors are $\alpha / \gamma = 0.435 + 0.039E_{\alpha}(MeV)$  and $\alpha / \gamma = 0.47 + 0.034E_{\alpha}(MeV)$ for \dma\ and \dmb, respectively, consistent with those reported in~\cite{ Bernabei:2008yh}.

For short-lived isotopes \iso{Po}{214} (\iso{U}{238}-chain) and \iso{Po}{212} (\iso{Th}{232}-chain), activities are estimated by Bi-Po events from data (see~\sect{sec:bkgd:bipo}). This analysis supports the contamination levels derived from the alpha spectrum.

The contamination levels of \iso{K}{40}\ and \iso{I}{129} were measured by using their continuous beta spectra with endpoints at 1311\kev\ and 154\kev, respectively, to match the simulation with the data.

%The absence of a 662\kev\ gamma line constrains the amount of \iso{Cs}{137} in the crystal.

%Due to the constraints of the deployment schedule, a thorough contamination measurement of the crystals was not performed prior to deployment. To derive the background levels in the crystals after deployment, simulation was matched to spectral features in the data (see~\tab{tab:nai}). 

%Background levels from the crystals were derived from a simulation matched to the spectral features described above (see~\tab{tab:nai}). Detector and background modeling were performed using the 4.9.5\_p01 release of the Geant4 software package~\cite{geant4, geant4b}.  The physics list was modeled off the ``rdecay02'' example code, which utilizes the Livermore processes for atomic deexcitation.

%%%%%%%%%%%%%%%%%%%%%%%%%%%%%%%%%%%%%%%%%%%%%%%%%%%
\subsection{Bi-Po events} \label{sec:bkgd:bipo}

\begin{figure}[!t]
	\includegraphics[width=\hw]{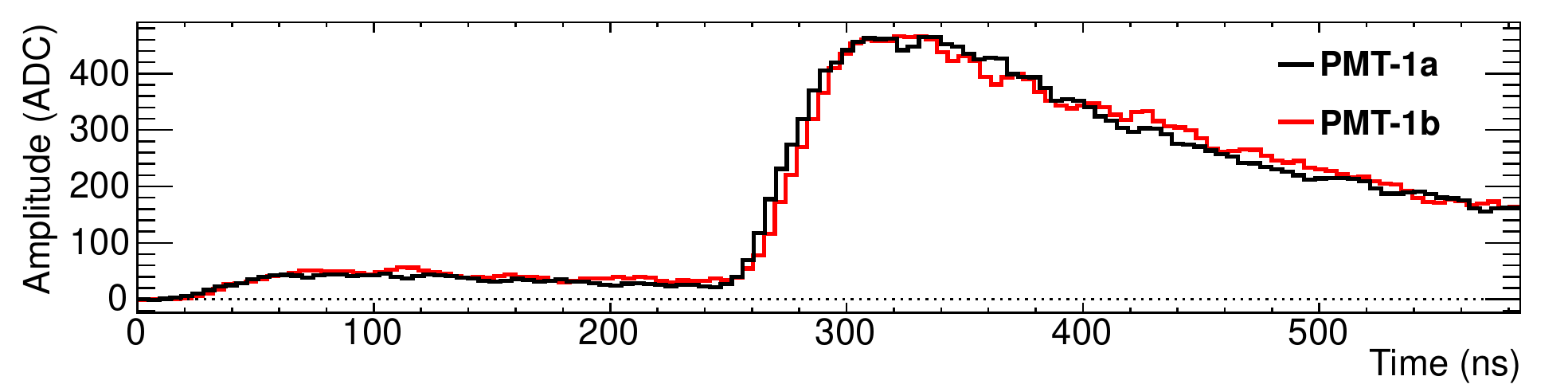}
	\caption{An example Bi-Po waveform recorded in the ATWD1 channels of PMT-1a (black) and PMT-1b (red). The first peak is the beta decay from \iso{Bi}{212}, and the second larger peak is from the alpha decay of \iso{Po}{212} ($t_{1/2}$\,=\,299\,ns).}
	\label{fig:bipo:wave} 
\end{figure}

For short-lived isotopes in a decay chain, it is possible to observe parent and daughter decays in a single recorded waveform (see~\fig{fig:bipo:wave}).  For the \aprox 600\,ns sample window of \dm, this most commonly occurs for the \iso{Th}{232}-chain decays \iso{Bi}{212}~$\rightarrow$~\iso{Po}{212}~$\rightarrow$~\iso{Pb}{208} (\iso{Po}{212} $t_{1/2}$\,=\,299\plm 2\,ns~\cite{nudat}).  A secondary contribution to the double-event population comes from the \iso{U}{238}-chain decays $\iso{Bi}{214} \rightarrow$ \iso{Po}{214} $\rightarrow$ \iso{Pb}{210} (\iso{Po}{214} $t_{1/2}$\,=\,164.3\plm 0.2\usec~\cite{nudat}).

Because the high-energy alpha event is delayed, a larger portion of the waveform tail is truncated.  This loss results in a correlated suppression of the calculated mean time and calibrated energy of these events (see~\fig{fig:tau1}).

\begin{figure}[!ht]
	\includegraphics[width=\hw]{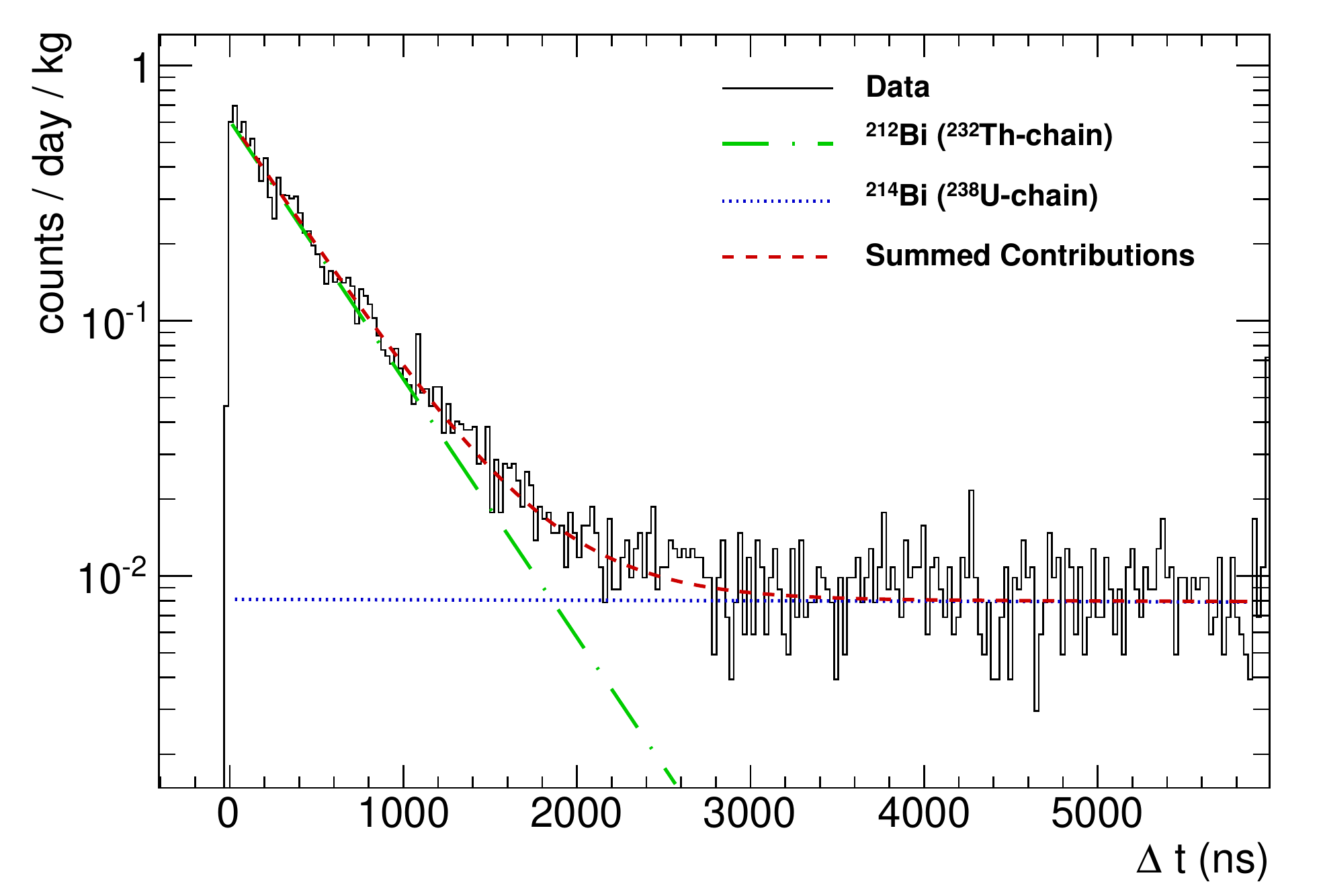}
	\caption{The time between gamma and subsequent alpha decay of Bi-Po events in PMT-1a. Exponential fit component from \iso{Bi}{212} decay (green, dot-dashed), flat continuum from \iso{Bi}{214} decay (blue, dotted), and composite fit (red, dashed) are overlain on data (black, solid).}
	\label{fig:bkgs:bipo} 
\end{figure}

The time between the two energy depositions in the waveform yields a distribution dominated by the exponential component due to \iso{Bi}{212} over the approximately flat continuum due to \iso{Bi}{214} (see~\fig{fig:bkgs:bipo}).  The exponential component of the fit yields a half-life of 298.6\plm 4.0\,ns, consistent with \iso{Po}{212} being the source of these events.

This analysis and the simulation comparison to the alpha peaks (see~\sect{sec:bkgd:nai}) yield similar results for the levels of contamination observed in the NaI crystals (see~\tab{tab:bipo}).

\begin{table}[!ht] 
\begin{ruledtabular}
\caption{Contamination levels of \iso{Bi}{212} (\iso{Th}{232}-chain) and \iso{Bi}{214} (\iso{U}{238}-chain) in the \dm\ NaI crystals derived from Bi-Po data analysis and simulation.  These values are consistent within the \aprox 30\per\ error of the simulation estimates.}
\label{tab:bipo}
	\begin{tabular}{c|cc|c}
	Isotope		& \multicolumn{2}{c|}{Bi-Po Analysis (\ubk)}	& Simulation	\\
				& \dma		& \dmb		& (\ubk)	\\
	\colrule
	\iso{Bi}{212}	& 176\plm 6	& 173\plm 6	& 160 \\
	\iso{Bi}{214}	& 930\plm 12	& 955\plm 12	& 900 \\
	\end{tabular}
\end{ruledtabular}
\end{table}

%%%%%%%%%%%%%%%%%%%%%%%%%%%%%%%%%%%%%%%%%%%%%%%%%%%
\section{Low Energy Region}
\label{sec:bkgd:noise}
\label{sec:bkgd:lowE}

%We observe four basic types of the digitized waveforms (as seen in \fig{fig:waveforms}): high energy scintillation, low energy scintillation, thin pulses, and electromagnetic interference (EMI). High energy (\gt 100\kev) scintillation events appear as a continuous waveform with the characteristic $\sim$300\,ns decay time of NaI(Tl). Low energy (\lt 100\kev) scintillation events appear as a photo-electron pulse train with more pulses, on average, near the beginning of the waveform. Thin pulses look like a single photo-electron except with a higher pulse amplitude compared to a photo-electron. EMI waveforms have a damped harmonic oscillator behavior.  The latter two pulse types can pollute the low energy signal region so must be removed before attempting to fit a background model to the data.

\begin{figure}[!ht]
	\includegraphics[width=\hw]{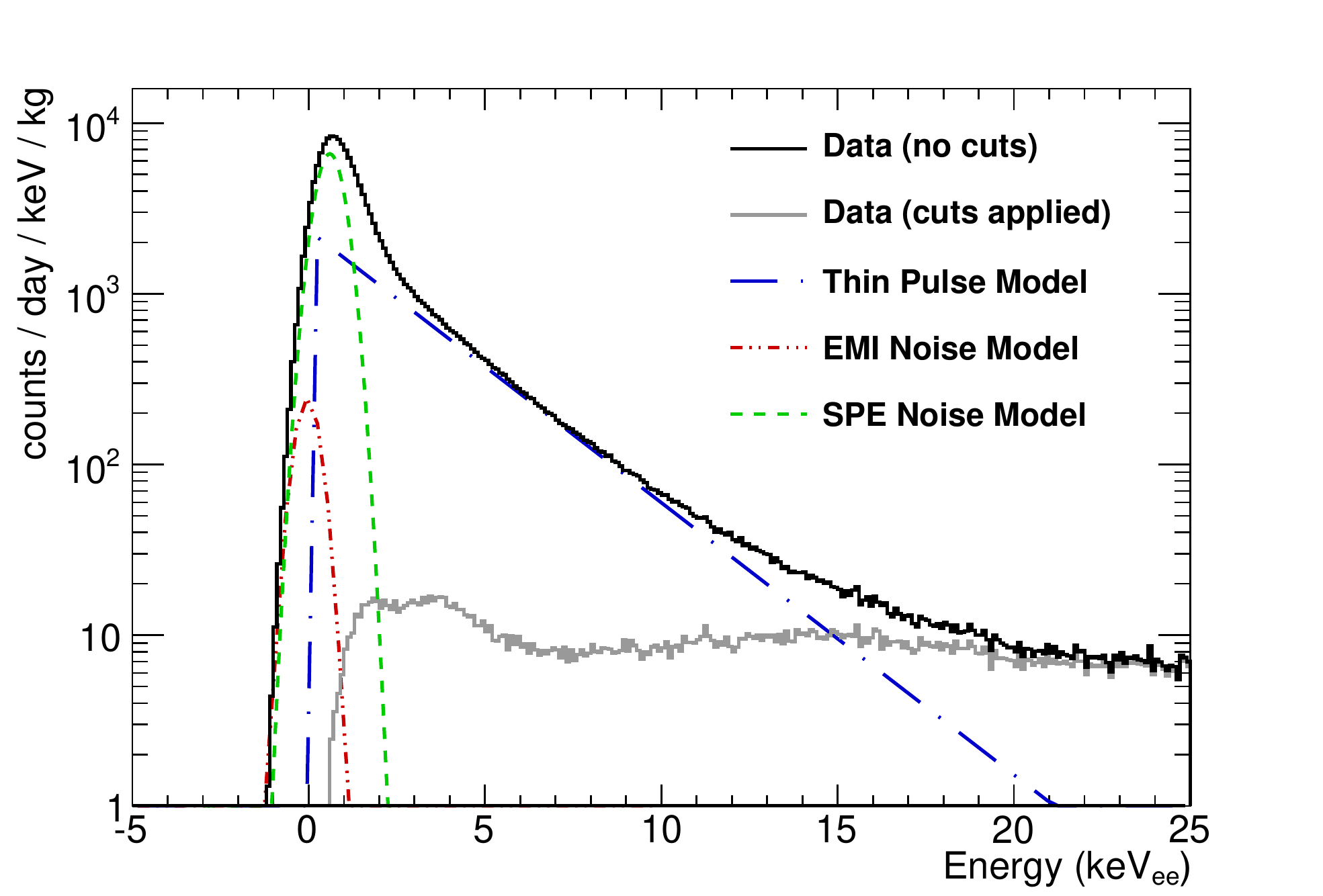}
	\caption{The low energy spectrum of \dma\ before (black, solid) and after (gray, solid) analysis cuts.  Models of the noise from thin pulses (blue, dot-dashed), electromagnetic interference (red, dot-dot-dashed), and correlated single photoelectrons (green, dashed) are shown.}
	\label{fig:cuts} 
\end{figure}

\begin{figure}[!ht]
	\subfigure[\ High-energy scintillation]{ %
		\includegraphics[width=\hw]{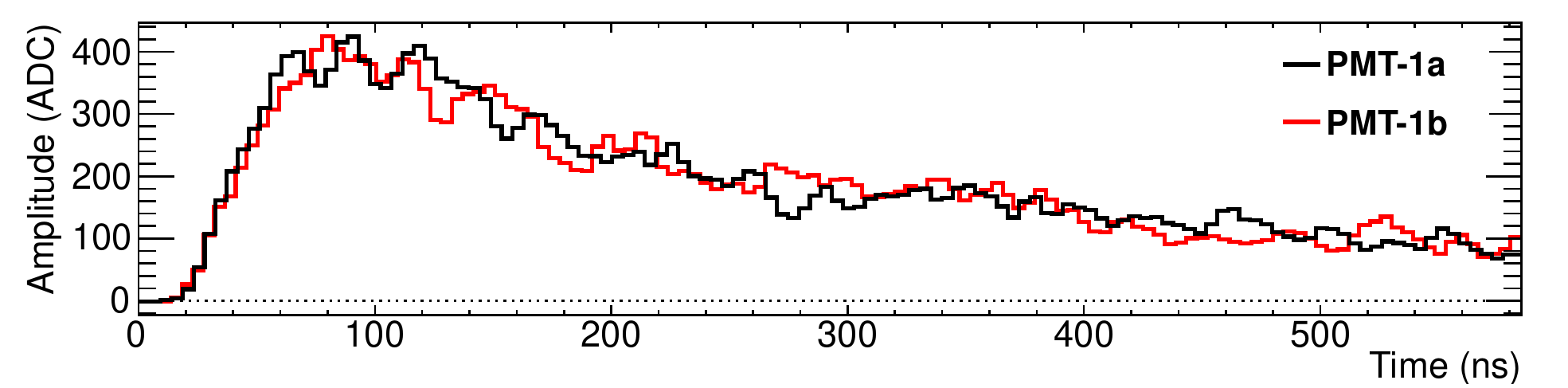} %
		\label{sub:scint} %
		} \\ %
	\subfigure[\ Low-energy scintillation]{ %
		\includegraphics[width=\hw]{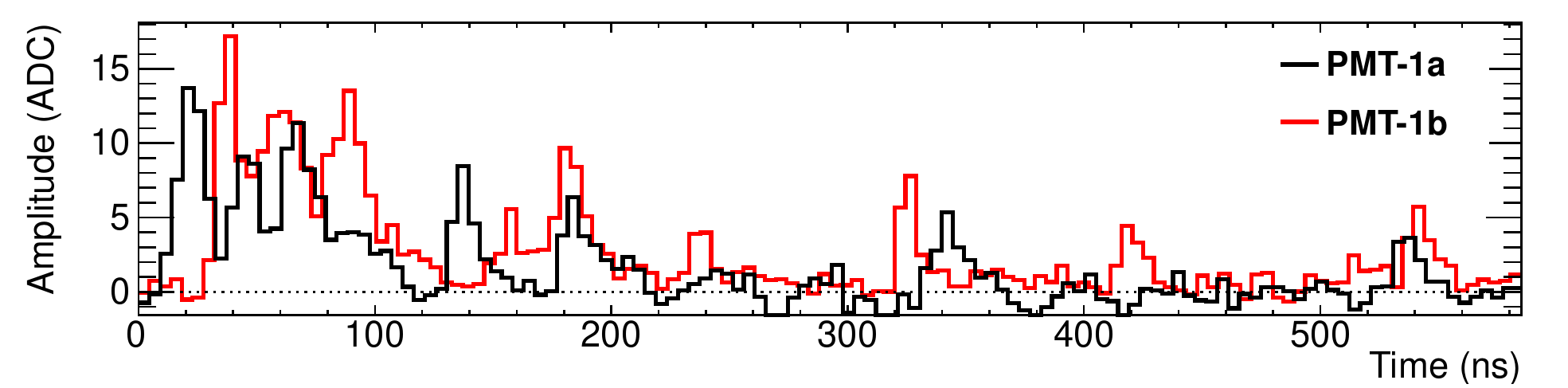} %
		\label{sub:spes} %
		} \\ %
	\subfigure[\ Thin pulses]{ %
		\includegraphics[width=\hw]{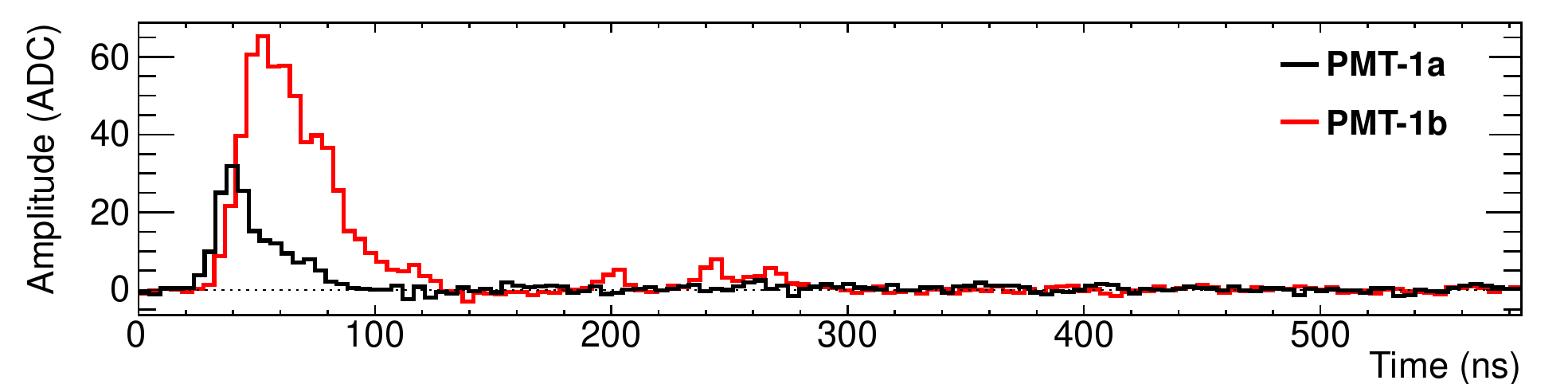} %
		\label{sub:thin} %
		} \\ %
	\subfigure[\ Electromagnetic interference]{ %
		\includegraphics[width=\hw]{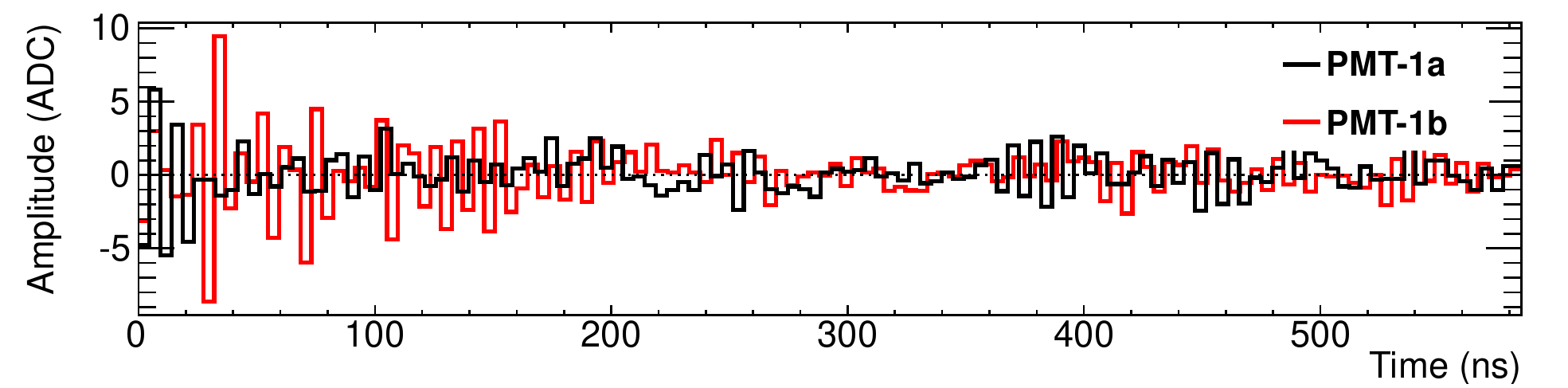} %
		\label{sub:emi} %
		} \\ %
	\caption{Example coincident waveforms of different event types recorded by PMT-1a (black) and PMT-1b (red). Waveforms (a) and (b) are signal events at high (700\kev) and low (7\kev) energies, respectively. Waveforms (c) and (d) are thin pulse (9\kev) and EMI (0.2\kev) noise events removed from raw data by cuts.}
	\label{fig:waveforms} 
\end{figure}

The energy spectrum below 25\kev\ is shown in~\fig{fig:cuts}. The raw spectrum (shown in solid black line) below 20\kev\ is dominated by three noise components: electromagnetic interference (EMI), thin pulses, and correlated single photoelectrons (SPE) between the two PMTs. The distinctive pulse shape of each event type is used to identify the noise; the spectrum after cuts is shown in gray. Thin pulses are the dominant noise source in the region of interest (2\,--\,20\kev), while EMI and SPE noise dominate below 1\kev. The waveforms associated with signal and these sources of noise are shown in~\fig{fig:waveforms}. 

EMI events occur when the mainboards are queried for their monitoring information (see~\sect{sec:dmice:elec}). These events are characterized by their oscillatory waveforms (see~\fig{sub:emi}), which can be distinguished from signal events by summing the square of the second derivative. This EMI cut parameter is
\begin{equation*}
	\sum_{n} \Big[ \left(ADC_{n+1} - ADC_{n}\right) - \left(ADC_{n} - ADC_{n-1} \right) \Big]^{2},
	\label{equ:emicut}
\end{equation*}
where ADC$_n$ is the $n$th time bin in the recorded waveform. Removal of EMI events is 99.98\per\ efficient while preserving 99.99\per\ of low energy non-EMI events. The calibrated energy of EMI events is centered at 0\kev; the highest energy observed is at 1.6\kev.  The monitoring query interval was increased from 2 to 60\,sec in February 2012, significantly reducing the number of EMI triggers.

Thin pulse events are characterized by tall, narrow waveforms and asymmetric energy deposition in the two PMTs (see~\fig{sub:thin}). Most of the light is collected within 100\,ns, a much faster decay time than that of NaI scintillation (\aprox 300\,ns in \dm\ data). Above 4\kev, thin pulse and scintillation event populations are well resolved using the ratio of the pulse area to the pulse height. The energy spectrum of thin pulses from 4\,--\,20\kev\ is exponential, and  is modeled to persist back to 0\kev\ (see~\fig{fig:cuts}). Thin pulses are suspected to originate from interactions in the quartz or PMT windows~\cite{Bernabei:2008yh, Amare:2014eea}.  Their timing characteristics are similar to noise pulses observed in other NaI detectors including those of DAMA/LIBRA~\cite{Bernabei:2008yh}. Understanding their stability and contribution to the signal region is critical for annual modulation studies.

%An important common feature of both correlated SPE and thin pulse noise is the lack of waveform features following a single deposition of light. In scintillation events, however, the photon arrivals are clustered at the beginning of the waveform but statistically spread through time. A cut on the number of peaks above a threshold exploits this characteristic difference between scintillation and noise events at the cost of imposing an effective threshold of \aprox 2\kev. 

Correlated SPE noise events form a Gaussian peak with a mean of 0.6\kev\ and a standard deviation of 0.4\kev\ (see green dashed line in \fig{fig:cuts}).  The shape matches the SPE spectrum observed in non-coincident characterization runs, but the rate is two orders of magnitude higher than expected from random coincidences between uncorrelated single photoelectrons from the two PMTs. The \aprox 100\,Hz SPE rates should produce \aprox 0.01\,Hz of accidental coincidences inside the 400\,ns coincidence window, but the observed rate of correlated SPE noise events is \aprox 1\,Hz.

Both thin pulse and SPE events are removed from the data set by cutting on the number of ``peaks'' (local maxima) observed in the recorded waveform.  Low-energy scintillation events (see~\fig{sub:spes}) are composed of multiple photoelectrons (5.9 and 4.3\,pe/keV for \dma\ and \dmb, respectively; see~\sect{sec:perf:res}), whereas thin pulse and SPE events typically only contain one peak.  The cut efficiency can be analyzed down to 4\,keV using pulse shapes, at which energy the signal passing efficiency is 60\% with signal to noise ratio of 16. The cut efficiency below 4\kev\ is a topic of ongoing investigation.

\begin{figure}[!ht]
	\includegraphics[width=\hw]{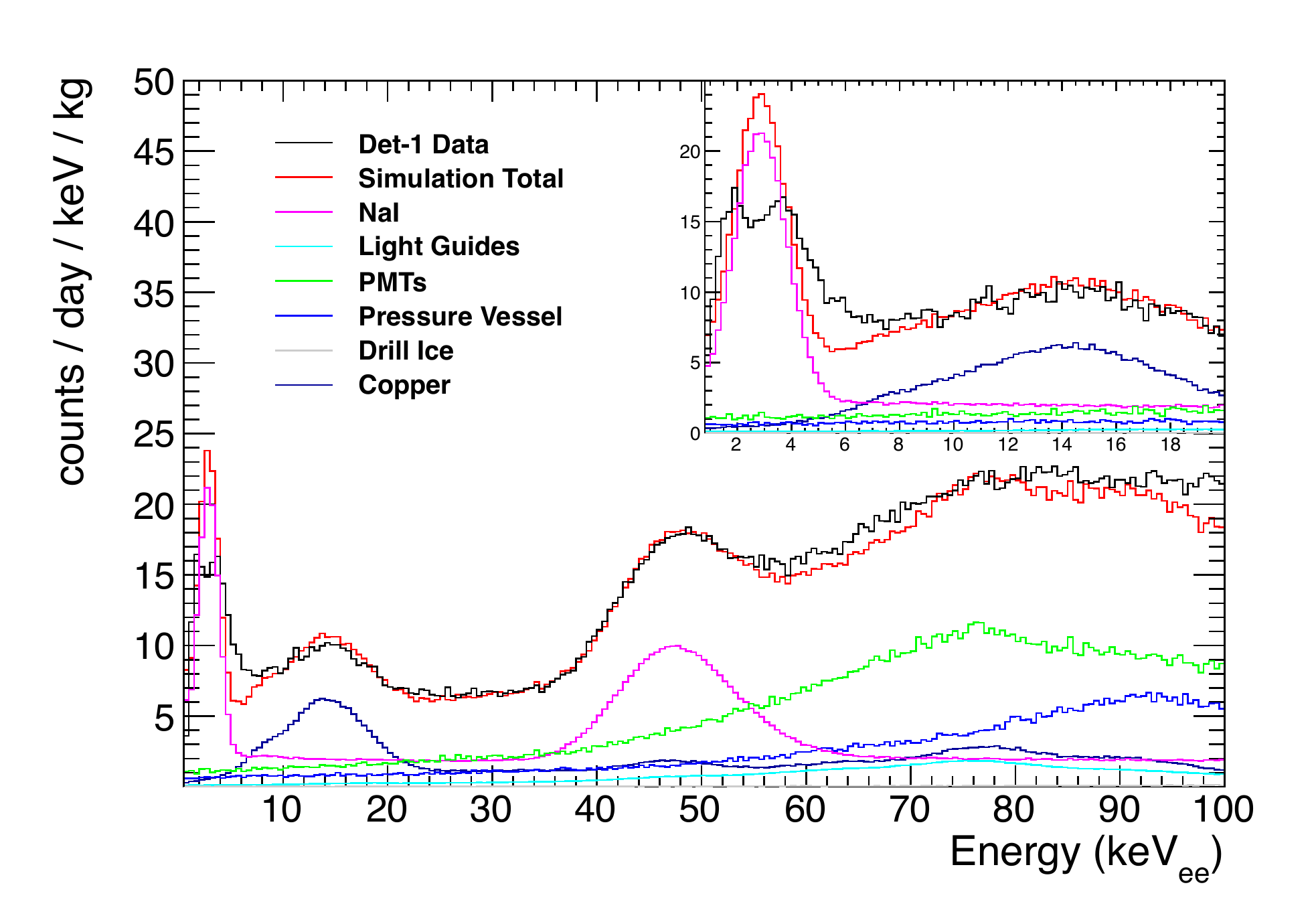}
	\caption{The  energy spectrum of \dm\ \dma\ (black) and simulation total (red). Simulated contributions from nearby detector components are also shown. The spectrum below 20\kev\ consists primarily of the 3\kev\ peak from \iso{K}{40} and a 14\kev\ peak from \iso{U}{238}-chain decays on the surface of the copper encapsulation.
	The background observed between 6.5\,--\,8\,\kev\ is 7.9\plm 0.4\dru.} 
	\label{fig:lowE:sim} 
\end{figure}

Several low energy features are visible in the data after applying cuts: a broad peak around 14\kev, a peak at 3\kev, and remnant noise below 2\kev.  These features are reproduced by \textsc{Geant4} simulations (see~\fig{fig:lowE:sim}). The broad peak at 14\kev\ can be attributed mostly to surface contamination of the \iso{U}{238}-chain in the copper encapsulation (40\,mBq from the inner 10\,$\upmu$m); surface contamination of this type has been observed previously in other NaI(Tl) experiments~\cite{Cebrian201260}. The 3\kev\ peak is from Auger electrons and X-rays from \iso{K}{40} decays in the crystal. The flat background of the crystal out to 30\kev\ is dominated by contributions from \iso{Pb}{210} (\iso{U}{238}-chain) and \iso{K}{40}.

A background of 7.9\plm 0.4\dru\ is observed between 6.5\,--\,8\kev, below which the spectrum is dominated by the \iso{K}{40} peak.  DAMA/LIBRA reports a single-hit spectrum with 0.9\dru\ background level after application of a multi-crystal anti-coincidence veto~\cite{Bernabei:2008yh}. Recently grown crystals being tested by ANAIS~\cite{Amare:2013lca} and KIMS~\cite{Kim:2014toa} have achieved a background level of 3\,--\,4\dru\ after cuts.

\section{Conclusions}

\dm\ was successfully constructed and deployed in December 2010.  The data presented here from July 2011 -- June 2013 demonstrate the excellent environmental conditions, including temperature stability with daily RMS of 0.02\deg. A livetime of \aprox 99\per\ was achieved by the detectors during this period.

It has been shown for the first time that low-background scintillator detectors can be remotely calibrated and operated under the ice at the South Pole.  This has promising implications for future dark matter searches in the Southern Hemisphere.

The background observed in \dm\ is in good agreement with simulation of expected contaminants levels in detector components; contributions from backgrounds in the environmental ice have been shown to be negligible. Below 20\kev, the spectrum is dominated by contributions from the crystals and encapsulation; in the 6.5\,--\,8.0\kev\ region, a rate of 7.9\plm 0.4\dru\ is observed, consistent with expected backgrounds. The energy region of interest between 2 and 6\kev\ is dominated by the events from \iso{K}{40} peak at 3\kev. An active research and development program toward achieving lower contaminant levels, particularly of \iso{K}{40} and \iso{U}{238}-chain, is currently underway. The backgrounds observed in new DM-Ice crystals as well as DM-Ice17 analyses of the low-energy region, muon coincidence with IceCube, and cosmogenic activation will be discussed in separate papers.

%Studies of the low-energy region, including a modulation analysis, and of muon coincidence with IceCube will be discussed in separate papers. R\&D paper
% R\&D are under way for the construction of lower background detectors that are able to directly test DAMA's assertion that their observed annual modulation is due to dark matter.%
\begin{acknowledgments}

We thank the IceCube Collaboration and construction team for their support in the successful deployment of the detector and on-going detector monitoring and data management. We thank the support from SNOLAB for their efforts on the low-background measurements, and STFC and the Boulby mine company CPL for support of low-background measurements of NaI.  This work was supported by the NSF Grants No.~PLR-1046816 and No.~PHY-1151795, Wisconsin IceCube Particle Astrophysics Center, the Wisconsin Alumni Research Foundation, Yale University, the Natural Sciences and Engineering Research Council of Canada, and Fermilab, operated by Fermi Research Alliance, LLC under Contract No.~DE-AC02-07CH11359 with the United States Department of Energy. W.~P.~and A.~H. are supported by the DOE/NNSA Stewardship Science Graduate Fellowship (Grant No.~DE-FC52-08NA28752) and NSF Graduate Research Fellowship (Grant No.~DGE-1256259), respectively. 

\end{acknowledgments}%

%%  BIBLIO %%%%%%%%%%%%%%%%%%%%%%%%%%%%%%%%%%%%%%%%%%%%%%%
%%%%%%%%%%%%%%%%%%%%%%%%%%%%%%%%%%%%%%%%%%%%%%%%%%%%%%%%%%%%
%%  References with bibTeX database:
%%%%%%%%%%%%%%%%%%%%%%%%%%%%%%%%%%%%%%%%%%%%%%%%%%%%%%%%%%%%
%%  natbib.sty is loaded by default. However, natbib options can be
%%  provided with \biboptions{...} command. Following options are
%%  valid:
%%%%%%%%%%%%%%%%%%%%%%%%%%%%%%%%%%%%%%%%%%%%%%%%%%%%%%%%%%%%
%%   round  -  round parentheses are used (default)
%%   square -  square brackets are used   [option]
%%   curly  -  curly braces are used      {option}
%%   angle  -  angle brackets are used    <option>
%%   semicolon  -  multiple citations separated by semi-colon
%%   colon  - same as semicolon, an earlier confusion
%%   comma  -  separated by comma
%%   numbers-  selects numerical citations
%%   super  -  numerical citations as superscripts
%%   sort   -  sorts multiple citations according to order in ref. list
%%   sort&compress   -  like sort, but also compresses numerical citations
%%   compress - compresses without sorting
%%%%%%%%%%%%%%%%%%%%%%%%%%%%%%%%%%%%%%%%%%%%%%%%%%%%%%%%%%%%

%\nocite{*}

\bibliographystyle{apsrev4-1}

\bibliography{%
bib/intro,%
bib/dmice,%
bib/else%
}

\end{document}